\documentstyle[12pt,psfig]{article}
\begin{document}
\begin{titlepage}
\vfill
{\large \bf
\begin{center}
Strong coupling constant to four loops in the analytic approach 
to QCD
\end{center}}
\vskip 1cm
%\normalsize
\renewcommand{\thefootnote}{\fnsymbol{footnote}}
\begin{center}
Aleksey I. Alekseev
\footnote{Electronic address: 
alekseev@mx.ihep.su}\\
{\small\it Institute for High Energy Physics,
142281 Protvino, Moscow Region, Russia}\\
\end{center}
\vskip 1.5cm
\begin{abstract}
The QCD analytic running coupling  $\alpha_{an}$
which has no nonphysical
singularities for all $Q^2>0$ 
is considered for the initial perturbation theory approximations up
to four loop order. The finiteness of the analytic coupling at
zero is shown to be a consequence of the  asymptotic freedom property
of the initial theory. 
The nonperturbative contributions to the analytic coupling are
extracted  explicitly.  For all $Q>\Lambda$ they are
represented in the form of
an expansion in  inverse powers of Euclidean momentum squared.
The effective method for a precise
calculation of the analytic running coupling is developed on the
basis of the stated expansion. 
The energy scale evolution of the
analytic running coupling for the one- to four-loop cases is studied
and the higher loop stability and low dependence on the quark
threshold matching conditions in comparison with the perturbative
running coupling were found. 
Normalizing the analytic running coupling at the scale of the rest
mass of the $Z$ boson
with the world average
value of the strong coupling constant,
$\alpha_{an}(M_Z^2)=$ $0.1181\pm 0.002$, 
one obtains as a result of the energy scale evolution of the
analytic running coupling   $\alpha_{an}(M_\tau^2)=$
0.2943$^{+0.0111}_{-0.0106}$ that is notably lower than the
estimations of the coupling strength available at the scale of the
mass of the $\tau$ lepton.
\end{abstract}
\vskip 1cm
PACS number(s): 12.38.Aw, 12.38.Lg
\vfill
\end{titlepage}
\section*{I. INTRODUCTION}
 The strong coupling constant $\alpha_s$ is the basic parameter of
Quantum Chromodynamics (QCD) and its determination appears to be one
of the most
important problems~\cite{Data,Bethke,Biebel}. 
The perturbation theory supplemented with the renormalization group
method works  effectively beyond the infrared region.
The nonphysical singularities of the 
perturbation theory arise in the infrared region of QCD and 
should be
canceled by the nonperturbative contributions.
%The so-called "analyticization procedure" is used in this approach.
The nonperturbative contributions arise quite naturally in an
analytic approach to QCD (for a review see, e.g.~\cite{SolShirTMF}).
%The main purpose of this 
%procedure is to remove nonphysical singularities from  approximate 
%(perturbative) expressions for the Green functions of QFT. 
The idea
of the procedure goes back to Refs.~\cite{Red,Bog} devoted to the
ghost pole problem in QFT. The foundation of the procedure is the
principle of summation of imaginary parts of the perturbation theory
terms. Then, the K\"allen -- Lehmann spectral representation  results
in the expressions without nonphysical singularities.
In recent papers~\cite{Shir,Shir1} it is suggested to solve the ghost
pole problem in QCD  demanding the 
$\alpha_s(Q^2)$  be analytic in $Q^2$ (to compare with the
dispersive approach~\cite{Doksh}). 
As a result, instead of the one-loop expression 
$\alpha^{(1)}_s(Q^2)=(4\pi/b_0)/\ln (Q^2/\Lambda^2)$
taking into account the leading logarithms and having the ghost
pole at $Q^2=\Lambda^2$ ($Q^2$ is the Euclidean momentum squared),
one obtains the  expression
\begin{equation}
\alpha^{(1)}_{an}(Q^2)=\frac{4\pi}{b_0}\left[\frac{1}{\ln(Q^2/
\Lambda^2)}+\frac{\Lambda^2}{\Lambda^2-Q^2}\right].
\label{1}
\end{equation}
Eq.~(\ref{1}) is an analytic function in the complex $Q^2$-plane
with a cut along the negative real semiaxis. The pole of the
perturbative running coupling at $Q^2=\Lambda^2$ is canceled by the
nonperturbative contribution 
$[\Lambda^2$ $\simeq \mu^2\exp
\{-4\pi/(b_0\alpha_{an}(\mu^2))\}$ at $\alpha_{an}(\mu^2)
\rightarrow 0]$ and the value $\alpha^{(1)}_{an}(0)=
4\pi/b_0$ appeared finite and independent of $\Lambda$. 
The  important feature of the "analyticization procedure" 
discovered~\cite{Shir,Shir1} is the stability property of the value
of the 
"analytically improved" running coupling constant at zero with 
respect to higher order corrections, 
$\alpha^{(1)}_{an}(0)=$	$\alpha^{(2)}_{an}(0)=$
$\alpha^{(3)}_{an}(0)$.
Though the derivative  of the analytic running coupling is infinite
at zero,
$\alpha_{an}(Q^2)$  turns out to be stable with 
respect to higher order corrections in the infrared region as a
whole.

The 1-loop order nonperturbative contribution in Eq.~(\ref{1}) can be
presented 
as convergent at $Q^2>\Lambda^2$ of constant signs series in the
inverse powers of the momentum squared. 
For a "standard" as well as for iterative 2-loop perturbative input
the nonperturbative contributions in analytic
running coupling are calculated explicitly in Ref.~\cite{A}.
In the ultraviolet region the nonperturbative contributions can also
be represented as a series in inverse powers of the momentum squared
with
different  coefficients of the expansion.
The nonperturbative 
contributions to $\alpha_{an}(Q^2)$ up to 3-loop order in
analytic approach to QCD are studied in Refs.~\cite{IHEP40},
\cite{YadFiz}. To handle the singularities originating from the
perturbative input  
the method which is more general than that of
Ref.~\cite{A} was developed. In Ref.~\cite{YadFiz} the momentum
dependence of $\alpha_{an}$ and its perturbative and nonperturbative
components in the infrared region are analysed. For the standard
perturbative input  the higher loop stability and low sensitivity
with respect to the $c$ quark threshold matching conditions were
found for $\alpha_{an}$.
In Ref.~\cite{JofP} the nonperturbative contributions for the 4-loop
case are considered briefly.

In this paper the momentum dependence of the analytic
running coupling up to 4-loop order is studied. In parallel, the
behavior of the
perturbative component is given for convenience of comparison.
In Section~2 we generalize slightly the standard four-loop solution
for $\alpha_s$, find and study the spectral density for $\alpha_{an}$
and then we prove the important property $\alpha_{an}(0)=$
$4\pi/b_0$. 
In Section~3 we extract from $\alpha_{an}$ 
the initial perturbative contribution $\alpha^{pt}$ and find in an
explicit form the
nonperturbative contribution $\alpha^{npt}_{an}$. We  develop a
technique of integration in the vicinity of  severe singularities 
of the perturbation theory in the infrared region and represent
$\alpha^{npt}_{an}$
in the form of a finite limits integral.
In Section~4 the power series representation for $\alpha^{npt}_{an}$
at $Q>\Lambda$ is obtained. 
In Section~5 we study the momentum behavior of $\alpha_{an}$. 
We consider it instructive to normalize $\alpha_{an}$ and
$\alpha^{pt}$ at $M_Z$ and
then compare their behavior to estimate the nonperturbative
contributions at all momenta.
We consider two methods of  matching of the solutions with different
numbers $n_f$ of active quark flavors.
Finally in Section~6 we give our conclusions.
In the Appendix we give the explicit formulas which allow one to
simplify
the
integration in the
vicinity of the  singularities of the standard perturbation theory
input.

\section*{II. FROM THE RUNNING COUPLING TO THE ANALYTIC RUNNING
COUPLING}
The behavior of the QCD running coupling $\alpha_s(Q^2)$
is defined by the renormalization group equation
\begin{equation}
Q^2\frac{\partial\alpha_s(Q^2)}{\partial Q^2}=
\beta(\alpha_s)=\beta_0\alpha_s^2+\beta_1\alpha_s^3+\beta_2\alpha_s^4
+\beta_3\alpha_s^5+O(\alpha_s^6),
\label{3}
\end{equation}
where the coefficients~\cite{Gross} --- \cite{Larin97}
\begin{eqnarray}
\beta_0=&-&\frac{1}{4\pi}b_0, \,\,\, b_0=11-\frac{2}{3}n_f,
\nonumber\\
\beta_1=&-&\frac{1}{8\pi^2}b_1, \,\,\, b_1=51-\frac{19}{3}n_f,
\nonumber\\
\beta_2=&-&\frac{1}{128\pi^3}b_2, \,\,\, b_2=
2857-\frac{5033}{9}n_f +\frac{325}{27}n_f^2,
\nonumber\\
\beta_3=&-&\frac{1}{256\pi^4}b_3, \,\,\, b_3=\frac{149753}{6}+
3564\zeta_3
\nonumber\\
&-&\left(\frac{1078361}{162}+\frac{6508}{27}\zeta_3\right)n_f+
\left(\frac{50065}{162}+\frac{6472}{81}\zeta_3\right)n_f^2+
\frac{1093}{729}n_f^3.
\label{4}
\end{eqnarray}
Here $n_f$ is the number of active quark
flavors and $\zeta$ is the Riemann zeta-function, $\zeta_3=\zeta(3)=
1.202056903...\,$. 
The first two coefficients $\beta_0$, $\beta_1$ do not depend on
the renormalization scheme choice. The next coefficients do depend on
it. Calculated within the
$\overline{MS}$-scheme in an arbitrary covariant gauge for the gluon
field they appeared to be independent of the gauge parameter choice.
Values of the coefficients~(\ref{4}) are given  in 
Table~\ref{t1}. All these  coefficients are small enough and decrease
in absolute value with
$n_f$ increasing. All the coefficients
are negative except $\beta_2$ at $n_f=6$.
\begin{table}[tbp]
\caption{$n_f$ dependence of $\overline{MS}$ values of $\beta_i$ 
($i= \overline{0,3}$), $b$, $\kappa$, $\bar\kappa$.}
\label{t1}
\begin{center}
\begin{tabular}{c| c c c c c c c}\hline \hline
     $n_f$& $\beta_0$  &$\beta_1$  & $\beta_2$ 
	 & $\beta_3$& $b$ &$\kappa$ &
	 $\bar\kappa$\\ \hline
0
&    -0.87535&    -0.64592&    -0.71986&    -1.17269&     0.84298&    
 0.51033&    -1.16716\\
1
&    -0.82230&    -0.56571&    -0.58199&    -0.91043&     0.83663&     
0.49541&    -1.20019\\
2
&    -0.76925&    -0.48550&    -0.45019&    -0.68103&     0.82045&     
0.46922&    -1.26081\\
3
&    -0.71620&    -0.40528&    -0.32445&    -0.48484&     0.79012&    
0.41467&    -1.36791\\
4
&    -0.66315&    -0.32507&    -0.20477&    -0.32222&     0.73920&     
0.28506&    -1.56255\\
5
&    -0.61009&    -0.24486&    -0.09116&    -0.19354&     0.65784&    
-0.07234&    -1.95343\\
6
&    -0.55704&    -0.16465&     0.01638&    -0.09914&     0.53061&    
-1.33654&    -2.94623\\
	 \hline \hline
\end{tabular}
\end{center}
\end{table}

The integration of Eq.~(\ref{3}) yields
\begin{eqnarray}
\frac{1}{\alpha_s(Q^2)}&+&\frac{\beta_1}{\beta_0}\ln\alpha_s(Q^2)+
\frac{1}{\beta_0^2}\left(\beta_0\beta_2-\beta_1^2\right)
\alpha_s(Q^2)+\frac{1}{2\beta_0^3}\left(\beta_1^3-
2\beta_0\beta_1\beta_2+\beta_0^2\beta_3\right)\alpha_s^2(Q^2)
\nonumber\\
&+&O(\alpha_s^3(Q^2))=-\beta_0\ln(Q^2/\Lambda^2)+\bar{C}.
\label{5}
\end{eqnarray}
The integration constant is represented here as a combination
of two  constants $\Lambda$ and $\bar{C}$.
Dimensional constant $\Lambda$ is a parameter which defines the 
scale of $Q$ and is used  for developing the iteration
procedure. 
Iteratively solving Eq.~(\ref{5}) for $\alpha_s(Q^2)$ at
$L=\ln(Q^2/\Lambda^2)\rightarrow\infty$ we obtain
\begin{eqnarray}
\frac{1}{\alpha_s(Q^2)}=&-&\beta_0 L+\frac{\beta_1}{\beta_0}
\left(\ln L+C\right)-\frac{\beta_1^2}{\beta_0^3 L}\left(\ln L+C+1-
\frac{\beta_0\beta_2}{\beta_1^2}\right)\nonumber\\
&-&\frac{\beta_1^3}{2\beta_0^5 L^2}\left[\left(\ln L+C\right)^2
-\frac{2\beta_0\beta_2}{\beta_1^2}\left(\ln L+C\right)-1+
\frac{\beta_0^2\beta_3}{\beta_1^3}\right]+O\left(\frac{1}{L^3}\right)
,\label{6}
\end{eqnarray}
where $C=\ln(-\beta_0)+(\beta_0/\beta_1)\bar{C}$.
Inverting Eq.~(\ref{6}) one obtains
$$
\alpha_s(Q^2)=-\frac{1}{\beta_0 L}\left\{1+\frac{\beta_1}{\beta_0
^2L}\left(\ln L+C\right)+\frac{\beta_1^2}{\beta_0^4L^2}
\left[\left(\ln L+C\right)^2-\left(\ln L+C\right)-1+\frac{\beta_0
\beta_2}{\beta_1^2}\right]\right.
$$
\begin{equation}
+\left.\frac{\beta_1^3}{\beta_0^6L^3}\left[\left(\ln L+C\right)^3
-\frac{5}{2}\left(\ln L+C\right)^2-\left(2-\frac{3\beta_0\beta_2}
{\beta_1^2}\right)\left(\ln L+C\right)+\frac{1}{2}-\frac{\beta_0^2
\beta_3}{2\beta_1^3}\right]+O\left(\frac{1}{L^4}\right)\right\}.
\label{7}
\end{equation}
 Within the
conventional definition of $\Lambda$ as 
$\Lambda_{\overline{MS}}$ \cite{Bardeen78} one chooses 
$C=0$. At that the functional form of the approximate solution for
$\alpha_s(Q^2)$ turns out to be somewhat simpler, but it
requires distinct $\Lambda_{\overline{MS}}$ for different $n_f$.
With this choice, Eq.~(\ref{7})  at the three loop level corresponds
to the standard  solution  written in the form of the expansion 
in inverse powers of logarithms~\cite{Data},
and at the four loop level it corresponds to~\cite{Chet97}.
We shall deal with nonzero $C$ since this freedom  can be useful
for an optimization of the finite order perturbation calculations.
Moreover, in the presence of the  $n_f$-dependent constant $C$ it is
possible to construct matched solution of Eq.~(\ref{3}) with
universal  $n_f$ independent constant $\Lambda$ \cite{Marciano}.

Let us introduce the function
$a(x)=(b_0/4\pi)\alpha_s(Q^2)$, where $x=Q^2/\Lambda^2$.
Then instead of~(\ref{7}) one can write
\begin{eqnarray}
a(x)&=&\frac{1}{\ln x}-b\frac{\ln(\ln x)+C}{\ln^2x}
+b^2\left[\frac{\left(\ln(\ln x)+C\right)^2}{\ln^3x}-\frac{\ln(\ln x)
+C}{\ln^3x}
+\frac{\kappa}{\ln ^3x}\right]
\nonumber\\
&-&b^3\left[\frac{\left(\ln(\ln x)+C\right)^3}{\ln^4 x}-\frac{5}{2}
\frac{\left(\ln(\ln x)+C\right)^2}{\ln^4x}+(3\kappa+1)\frac{
\ln(\ln x)+C}{\ln^4x}+\frac{\bar\kappa}{\ln^4x}\right].
\label{9}
\end{eqnarray}
where the coefficients $b$, $\kappa$, and $\bar\kappa$ are equal to
\begin{eqnarray}
b&=&-\frac{\beta_1}{\beta^2_0}=\frac{2b_1}{b^2_0},\nonumber\\
\kappa&=&-1+\frac{\beta_0 \beta_2}{\beta_1^2}=-1+\frac{b_0 b_2}
{8b^2_1},\nonumber\\
\bar\kappa&=&\frac{1}{2}-\frac{\beta_0^2\beta_3}{2\beta_1^3}=
\frac{1}{2}-\frac{b_0^2b_3}{16b_1^3}.
\label{10}
\end{eqnarray}
The values of parameters $b$, $\kappa$, and $\bar\kappa$ of
Eq.~(\ref{9}) for different $n_f$ are given in Table~\ref{t1}.
At $x\simeq 1$   the perturbative running coupling is singular.
At large $x$  the 1-loop term of Eq.~(\ref{9}) 
defines the ultraviolet
behavior of $a(x)$. However,  
for small $x$ the behavior of the running coupling depends on the 
approximation we adopt
and at
$x=1$ there are singularities  of a different analytical structure.
Namely, at $x\simeq 1$ the leading singularities are
\begin{eqnarray}
a^{(1)}(x)&\simeq&\frac{1}{x-1},\,\,\,
a^{(2)}(x)\simeq-\frac{b}{(x-1)^2}\ln (x-1),\nonumber\\
a^{(3)}(x)&\simeq&\frac{b^2}{(x-1)^3}\ln^2 (x-1),\,\,\,
a^{(4)}(x)\simeq-\frac{b^3}{(x-1)^4}\ln^3 (x-1).
\label{11}
\end{eqnarray}
From Eqs.~(\ref{11}) we know the leading behavior at  $x\simeq 1$ of
the additional terms which
should cancel the perturbative singularities. But in principle it
gives no information on their behavior at large $x$. 
The analytic approach  removes
all these nonphysical singularities in a regular way.

The analytic running coupling is 
obtained  by the integral
representation
\begin{equation}
a_{an}(x)=\frac{1}{\pi}\int\limits_0^\infty \frac{d\sigma}{x+\sigma}
\rho(\sigma),
\label{2}
\end{equation}
where the spectral density $\rho(\sigma)={\rm Im}
a_{an}(-\sigma-i0)$.
According to the analytic approach to QCD we adopt that
${\rm Im} a_{an}(-\sigma-i0)={\rm Im} a(-\sigma-i0)$, where $a(x)$ is
the perturbative running coupling.
It is clear that dispersively-modified coupling of  
form~(\ref{2}) has
analytical structure which is consistent with causality.

By making the analytic continuation of  Eq.~(\ref{9}) into 
the Minkowski
space $x=-\sigma-i0$, one  obtains
$$
a(-\sigma-i0)=\frac{1}{\ln \sigma-i\pi}-b\frac{
\ln\left(\ln\sigma-i\pi\right)+C}{(\ln \sigma
-i\pi)^2}
+b^2\left\{\frac{[\ln(\ln\sigma-i\pi)+C]^2}{(\ln\sigma-i\pi)^3}
\right.
$$
$$
-\left.\frac{\ln(\ln\sigma-i\pi)+C}{(\ln\sigma-i\pi)^3}+
\frac{\kappa}{(\ln\sigma-i\pi)^3}\right\}
-b^3\left\{\frac{[\ln(\ln \sigma-i\pi)+C]^3}{(\ln\sigma-i\pi)^4}
-\frac{5}{2}
\frac{[\ln(\ln \sigma-i\pi)+C]^2}{(\ln\sigma-i\pi)^4}
\right.
$$
\begin{equation}
+\left.(3\kappa+1)\frac{
\ln(\ln \sigma-i\pi)+C}{(\ln\sigma-i\pi)^4}+\frac{\bar\kappa}
{(\ln\sigma-i\pi)^4}\right\}.
\label{12}
\end{equation}
Taking an imaginary part of Eq.~(\ref{12}) we find the spectral
density
\begin{equation}
\rho(\sigma)=\rho^{(1)}(\sigma)+\Delta\rho^{(2)}(\sigma)+
\Delta\rho^{(3)}(\sigma)+\Delta\rho^{(4)}(\sigma),
\label{ro}
\end{equation}
where
\begin{equation}
\rho^{(1)}(\sigma)=
\frac{\pi}{t^2+\pi^2},
\label{ro1}
\end{equation}
\begin{equation}
\Delta\rho^{(2)}(\sigma)=
-\frac{b}{(t^2+\pi^2)^2}\left[
2\pi tF_1(t)-\left(t^2-\pi^2\right)F_2(t)\right],
\label{ro2}
\end{equation}
$$
\Delta\rho^{(3)}(\sigma)=\frac{b^2}{(t^2+\pi^2)^3}\left[\pi
\left(3t^2-\pi^2\right)\left(
F_1^2(t)-F_2^2(t)\right)-
2t\left(t^2-3\pi^2\right)F_1(t)F_2(t)\right.
$$
\begin{equation}
-\pi\left(3t^2-\pi^2\right)F_1(t)
\left.+t\left(t^2-3\pi^2\right)F_2(t)
+\pi\kappa\left(3t^2-\pi^2\right)\right],
\label{ro3}
\end{equation}
$$
\Delta\rho^{(4)}(\sigma)=-\frac{b^3}{(t^2+\pi^2)^4}\left[\left(t^4-
6\pi^2t^2+\pi^4\right)\left(F_2^3(t)-3F_1^2(t)F_2(t)
\right)+4\pi t\left(t^2-
\pi^2\right)\left(F_1^3(t)\right.\right.
$$
$$
\left. -3F_1(t)F_2^2(t)\right)
-10\pi t\left(t^2-\pi^2\right)\left(F_1^2(t)-F_2^2(t)
\right)+5\left(t^4-
6\pi^2t^2+\pi^4\right)F_1(t)F_2(t)
$$
\begin{equation}
+4\pi \left(1+3\kappa\right)t\left(
t^2-\pi^2\right)F_1(t)
-\left.\left(1+3\kappa\right)\left(t^4-6\pi^2t^2+\pi^4\right)F_2(t)+
4\pi \bar\kappa t\left(t^2-\pi^2\right)\right].
\label{ro4}
\end{equation}
Here $t=\ln(\sigma)$, 
 \begin{equation}
F_1(t)\equiv\frac{1}{2}\ln(t^2+\pi^2)+C,\,\,\,
F_2(t)\equiv\arccos\frac{t}{\sqrt{t^2+\pi^2}},
\label{ro5}
\end{equation}
$\rho^{(1)}(\sigma)$ is the 1-loop spectral density and
$\Delta\rho^{(l)}(\sigma)$ are higher loop corrections to the
spectral
density. With Eqs.~(\ref{2}), (\ref{ro}) --- (\ref{ro5}) the analytic
running
coupling can be studied, e.g. by  numerical methods.
For the 1 --- 4-loop cases the spectral density of the analytic
running coupling is shown in Fig.~\ref{fig2}.
For the curves in Fig.~\ref{fig2} and in the next  Fig.~\ref{fig1} 
the parameter values $C=0$, $n_f=3$ are chosen.
Beyond the 1-loop approximation   one
can see the higher loop stabilization of the
spectral density and in the region of $|t|>10$ it is practically the
same for the 2 --- 4-loop cases. Integrating the spectral density
numerically with  replacement of the infinite limits by finite cut
parameter $T$ leads to   
the relative error $\sim 1/T$, and at large $T$ it is important not
to lose the higher loop contributions. In Fig.~\ref{fig1} the higher
loop corrections to the spectral density are shown. In fact we deal
with rapidly oscillating functions and one needs special methods 
for a precise integration. E.g., for the 4-loop case it is difficult
to get a 2-percent accuracy for $\alpha_{an}(M^2_{\tau})$ using the
standard integration program DGAUSS.

We shall obtain another more effective method
for  precise calculation of $\alpha_{an}(Q^2)$ which is not
connected with the numerical integration.

Function $a(x)$ in Eq.~(\ref{9}) is regular and real for real $x>1$.
Thus, to find the spectral density $\rho(\sigma)$ we can use Schwarz 
reflection principle
$(a(x))^*=a(x^*)$ where $x$ is considered as a complex
variable. Then
\begin{equation}
\rho(\sigma)=\frac{1}{2i}\left( a(-\sigma-i0)-a(-\sigma+i0)\right).
\label{133}
\end{equation}

Let us introduce function $\Phi(z)$ of the form
\begin{eqnarray}
\Phi(z)&=&\frac{1}{z}-b\frac{\ln(z)+C}{z^2}
+b^2\left[\frac{\left(\ln(z)+C\right)^2}{z^3}-\frac{\ln(z)
+C}{z^3}
+\frac{\kappa}{z^3}\right]
\nonumber\\
&-&b^3\left[\frac{\left(\ln(z)+C\right)^3}{z^4 }-\frac{5}{2}
\frac{\left(\ln(z)+C\right)^2}{z^4}+(3\kappa+1)\frac{
\ln(z)+C}{z^4}+\frac{\bar\kappa}{z^4}\right].
\label{99}
\end{eqnarray}
To choose the main branch of the multivalued function~(\ref{99}) we 
cut the complex z-plane along the negative semiaxis. Then 
solution~(\ref{9}) can be written as 
$a(x)=\Phi(\ln x)$.  Function $a(x)$ is unambiguously defined in the
complex $x$-plain with two cuts along the real axis, physical cut
from minus infinity to zero and nonphysical one from zero to unity. 
Then
\begin{equation}
\rho(\sigma)=\frac{1}{2i}\left( \Phi(\ln\sigma-i\pi)-\Phi(\ln
\sigma+i\pi)\right).
\label{13}
\end{equation}
By the change of  variable of the form $\sigma=\exp (t)$,
the analytical expression
is derived from~(\ref{2}),  (\ref{13}) as follows:
\begin{equation}
a_{an}(x)=\frac{1}{2\pi i}\int\limits^\infty_{-\infty}
dt \, \frac{e^t}{x+e^t}\times
\left\{\Phi(t-i\pi)-\Phi(t+i\pi)
\right\}.
\label{14}
\end{equation}
Let us prove that $a_{an}(0)=1$. It follows from Eq.~(\ref{14}) that
$$
a_{an}(0)=\frac{1}{2\pi i}\int\limits^\infty_{-\infty}
dt \, 
\left\{\Phi(t-i\pi)-\Phi(t+i\pi)
\right\}
$$
\begin{equation}
=\frac{1}{2\pi i}\int\limits^\infty_{-\infty}
dt \, 
\left\{\left[\Phi(t-i\pi)-\frac{1}{t-i\pi}\right]-\left[\Phi(t+i\pi)-
\frac{1}{t+i\pi}\right]+\left[\frac{1}{t-i\pi}-\frac{1}{t+i\pi}\right
]
\right\}.
\label{144}
\end{equation}
For the first term in Eq.~(\ref{144}) we close the integration
contour in the lower half-plane of the complex variable $t$ by the
arch of the "infinite" radius without 
affecting the value of the integral. We can do it because
the integrand  multiplied by $t$ goes to
zero at $\mid t\mid\rightarrow \infty$.
There are no singularities
inside the contour, and thus we obtain a zero contribution from the
term
considered. For the second term  we close the integration contour
in the upper half-plane of the complex variable $t$ with the same
result. Therefore we have:
\begin{equation}
a_{an}(0)=\frac{1}{2\pi i}\int\limits^\infty_{-\infty}
dt \, 
\left[\frac{1}{t-i\pi}-\frac{1}{t+i\pi}\right]=1.
\label{145}
\end{equation}
For any finite loop order the expansion structure of the perturbative
solution in inverse powers of logarithms ensure the property of the
analytic coupling $a_{an}(0)=1$. The arguments are suitable for all
solutions $\Phi(z)$ as long as the singularities are situated at the
real axis of the 
complex $z$-plane, in particular for the iterative solutions of
Refs.~\cite{Shir,Shir1}.

\section*{III. EXTRACTION OF THE NONPERTURBATIVE CONTRIBUTIONS}

Let us see what the singularities of the integrand of~(\ref{14}) 
in the complex  $t$-plane are.
First of all the integrand has  simple
poles at  $t=\ln x\pm i\pi(1+2n)$,
$n=0,1,2,...$. All the residues of function  
$\exp(t)/(x+\exp(t))$ 
at these points are equal to unity.
Apart from these 
poles the integrand of~(\ref{14}) 
has at $t=\pm i\pi$  poles up to fourth order  and
logarithmic type branch points  which coincide with the poles
from the second order to the fourth order.
The initial integration contour in the complex $t$-plane and
singularities of the integrand of Eq.~(\ref{14}) are shown in
Fig.~3(a).
Let us cut the complex
$t$-plane in a  
standard way, $t=\pm i\pi-\lambda$, with $\lambda$ being the real
parameter varying from  $0$ to $\infty$.
Further on we append
the integration by the arch of the "infinite" radius without 
affecting the value of the integral, and close the integration
contour $C_1$ in the upper half-plane of the complex variable $t$
excluding the singularities at $t=i\pi$. In this case 
an additional contribution emerges due to the integration along
the sides of the cut and around the singularities at $t=i\pi$.
The corresponding contour we denote as $C_2$. The integration contour
$C_1$ is shown in Fig.~3(b), whereas the contour $C_2$ is shown in
Fig.~3(c).

Let us turn to the integration along the contour $C_1$.
For the integrand of Eq.~(\ref{14}), which we denote as
$F(t)$, the residues at $t=\ln x+i\pi(1+2n)$, $n=0,1,2,...$  are
as follows:
\begin{equation}
{\rm Res} F(t)\mid _{t=\ln x+i\pi(1+2n)}=
\Phi(\ln x+2\pi i n)-\Phi(\ln x+2\pi i(n+1)).
\label{200}
\end{equation}
By using the residue theorem  
one readily obtains  the  contribution $\Sigma(x)$ to the
integral~(\ref{14}) from the integration along the contour $C_1$
\begin{equation}
\Sigma(x)=\frac{1}{2\pi i}\int\limits_{C_1}F(t)\,dt= 
\sum\limits^\infty_{n=0}{\rm Res} F\left(t=\ln x+i\pi(1+2n)\right)
=\Phi(\ln x)=a(x).
\label{16}
\end{equation}
One can see that this contribution is exactly equal to the initial
Eq.~(\ref{9}). Therefore we call it a perturbative part of 
$a_{an}(x)$,  $a^{pt}(x)=\Sigma(x)$.
The remaining contribution of the integral along the contour $C_2$
can naturally be called a nonperturbative part of $a_{an}(x)$,
\begin{equation}
a_{an}(x)=a^{pt}(x)+a^{npt}_{an}(x).
\label{17}
\end{equation}
Let us turn to the calculation of  $a^{npt}_{an}(x)$. We can omit the
terms of the integrand in Eq.~(\ref{14}) which have no 
singularities at $t=i\pi$. Then we have
$$
a^{npt}_{an}(x)=\frac{1}{2\pi i}\int\limits_{C_2}
dt \, \frac{e^t}{x+e^t}\times
\left\{\frac{1}{t-i\pi}
-b\frac{\ln(t-i\pi)+C}
{(t-i\pi)^2}\right.
$$
$$
+b^2\left.\left[\frac{[\ln(t-i\pi)+C]^2}{(t-i\pi)^3}
-\frac{\ln(t-i\pi)+C}{(t-i\pi)^3}
+\frac{\kappa}{(t-i\pi)^3}
\right]\right.
$$
\begin{equation}
-\left.b^3\left[\frac{[\ln(t-i\pi)+C]^3}{(t-i\pi)^4}
-\frac{5}{2}\frac{[\ln(t-i\pi)+C]^2}{(t-i\pi)^4}
+(3\kappa+1)\frac{\ln(t-i\pi)+C}{(t-i\pi)^4}
+\frac{\bar\kappa}{(t-i\pi)^4}
\right]\right\}.
\label{18}
\end{equation}
Let us change the variable $t=z+i\pi$ and introduce the function
\begin{equation}
f(z)=\frac{1}{1-x\exp(-z)}.
\label{19}
\end{equation}
Then we can rewrite Eq.~(\ref{18}) in the form
$$
a^{npt}_{an}(x)=\frac{1}{2\pi i}\int\limits_{\tilde C} dz \, f(z)
\left\{\frac{1}{z}-b\left[\frac{\ln(z)}{z^2}+\frac{C}{z^2}\right]
+b^2\left[\frac{\ln^2(z)}{z^3}+\left(2C-1\right)\frac{\ln z}{z^3}+
\frac{\kappa-C+C^2}{z^3}\right]\right.
$$
\begin{equation}
-\left.b^3\left[\frac{\ln^3(z)}{z^4}+\left(3C-\frac{5}{2}\right)
\frac{\ln^2}{z^4}+(3C^2-5C+3\kappa+1)\frac{\ln z}{z^4}+
\frac{C^3-\frac{5}{2}C^2+(3\kappa+1)C+\bar\kappa}{z^4}\right]
\right\}.
\label{20}
\end{equation}
The cut in  the complex $z$-plane goes now from zero to $-\infty$.
Starting from $z=-\infty-i0$, the contour $\tilde C$ goes 
along the lower side of the cut, then
around the origin 
and then  further along the upper side of the cut  to  
$z=-\infty+i0$.  $x$ is considered here as a real  variable, $x>1$.
Then the contour $\tilde C$ can be chosen in such a way that it does
not
envelop  "superfluous" singularities, and the conditions used
in the Appendix for finding the corresponding integrals are
satisfied.
Function~(\ref{19}) with its derivatives 
\begin{eqnarray}
f'(z)&=&-\frac{x\exp(-z)}{(1-x\exp(-z))^2},\,\,\,
f''(z)=\frac{x\exp(-z)(1+x\exp(-z))}{(1-x\exp(-z))^3},\,\,\,
\nonumber\\
f'''(z)&=&-\frac{x\exp(-z)}{(1-x\exp(-z))^4}\left(1+4x\exp(-z)+
x^2\exp(-2z)\right),
\label{21}		 \\
f''''(z)&=&\frac{x\exp(-z)}{(1-x\exp(-z))^5}\left(1+11x\exp(-z)+
11x^2\exp(-2z)+x^3\exp(-3z)\right)\nonumber
\end{eqnarray}
decrease exponentially
at $z\rightarrow -\infty$. Therefore, we shall omit the boundary
terms
in formulas given in the Appendix\footnote{Using these formulas  with
$f=1$ one can make sure that $a(x=0)=1$. The boundary terms should be
considered in this case.}. 
Then, from Eq.~(\ref{20}) one can 
obtain
\begin{eqnarray}
a^{npt}_{an}(x)=&-&\frac{1}{2\pi i}\int\limits_{\tilde C}
dz \,\left\{
f'(z)\ln(z)-b\left[(1+C)\ln(z)+\frac{1}{2}\ln^2(z)\right]f''(z)
\right.\nonumber\\
&+&\frac{1}{2}b^2\left[\left(2+\kappa+2C+C^2\right)\ln(z)+
(1+C)\ln^2(z)+\frac{1}{3}\ln^3(z)\right]f'''(z)\nonumber\\
&-&\frac{1}{6}b^3\left[\left(6+\frac{11}{2}\kappa+\bar\kappa+
3(2+\kappa)
C+3C^2+C^3\right)\ln z
+\frac{3}{2}(2+\kappa+2C+C^2)\ln^2z\right.\nonumber\\
&+&\left.\left.(1+C)\ln^3z+\frac{1}{4}\ln^4z
\right]f''''(z)\right\}.
\label{22}
\end{eqnarray}
Taking into account that function $f(z)$ with its derivatives
is regular at real negative semiaxis of $z$
we can rewrite equation~(\ref{22}) in the form
\begin{eqnarray}
a^{npt}_{an}(x)=&-&\int\limits_{0}\limits^{-\infty} du \,\left\{
f'(u)\bar\Delta_1(u)-b\left[(1+C)\bar\Delta_1(u)+\frac{1}{2}\bar
\Delta_2(u)
\right]f''(u)
\right.\nonumber\\
&+&\frac{1}{2}b^2\left[\left(1+\kappa+(1+C)^2\right)\bar\Delta_1(u)+
(1+C)\bar\Delta_2(u)+\frac{1}{3}\bar\Delta_3(u)\right]f'''(u)
\nonumber\\
&-&\frac{1}{6}b^3\left[\left(2+\frac{5}{2}\kappa+\bar\kappa+3(1+C)
(1+\kappa) +(1+C)^3\right)\bar\Delta_1(u)\right.
\nonumber\\
&+&\frac{3}{2}\left(1+\kappa+(1+C)^2\right)\bar\Delta_2(u)
 +\left.\left.(1+C)\bar\Delta_3(u)+\frac{1}{4}\bar\Delta_4(u)
\right]f''''(u)\right\},
\label{23}
\end{eqnarray}
where $u$  is real,	$u<0$ and $\bar\Delta_i(u)$ are discontinuities
of 
the powers of the
logarithms
\begin{eqnarray}
\bar\Delta_1(u)&=&\frac{1}{2\pi i}\left(\ln(u+i0)-\ln(u-i0)\right)=1,
\nonumber \\
\bar\Delta_2(u)&=&\frac{1}{2\pi
i}\left(\ln^2(u+i0)-\ln^2(u-i0)\right)=
2\ln(-u),\nonumber\\
\bar\Delta_3(u)&=&\frac{1}{2\pi
i}\left(\ln^3(u+i0)-\ln^3(u-i0)\right)=
3\ln^2(-u)-\pi^2, \label{24} \\
\bar\Delta_4(u)&=&\frac{1}{2\pi
i}\left(\ln^4(u+i0)-\ln^4(u-i0)\right)=
4\ln^3(-u)-4\pi^2\ln(-u).  \nonumber
\end{eqnarray}
Let us introduce the variable $\sigma=\exp(u)$. From Eqs.~(\ref{21}), 
(\ref{23}), (\ref{24}) we obtain
$$
a^{npt}_{an}(x)=-x\int\limits_{0}\limits^{1} d\sigma \,\left\{
\frac{1}{(x-\sigma)^2}-b\biggl[1+C+\ln(-\ln\sigma)
\biggr]\frac{x+\sigma}{(x-\sigma)^3}
\right.
$$
$$
+\frac{1}{2}b^2\left[1-\frac{\pi^2}{3}+\kappa+(1+C)^2+
2(1+C)\ln(-\ln\sigma)+\ln^2(-\ln\sigma)\right]\frac{x^2+4x\sigma+
\sigma^2}{(x-\sigma)^4}
$$
$$
-\frac{1}{6}b^3\left[2+\frac{5}{2}\kappa+\bar\kappa+3(1+C)
(1-\frac{\pi^2}{3}+\kappa) +(1+C)^3
+3\left(1-\frac{\pi^2}{3}+\kappa+(1+C)^2\right)\right.
$$
\begin{equation}
\times
\ln(-\ln\sigma)
+\left.3(1+C)\ln^2(-\ln\sigma)+\ln^3(-\ln\sigma)
\biggr]\frac{x^3+11x^2\sigma+11x\sigma^2+\sigma^3}{(x-\sigma)^5}
\right\}.
\label{25}
\end{equation}
Integrating the terms of Eq.~(\ref{25}) independent of logarithms
one can obtain
$$
a^{npt}_{an}(x)=-\frac{1}{x-1}+b\left\{\frac{(1+C)x}{(x-1)^2}+
x\int\limits_{0}\limits^{1} d\sigma \,\ln(-\ln\sigma)
\frac{x+\sigma}{(x-\sigma)^3}\right\}
$$
$$
-\frac{1}{2}b^2\left\{\left[1-\frac{\pi^2}{3}+\kappa+(1+C)^2\right]
\frac{x(x+1)}{(x-1)^3}
+x\int\limits_{0}\limits^{1} d\sigma \,\biggl[
2(1+C)\ln(-\ln\sigma)+\ln^2(-\ln\sigma)\biggr]\right.
$$
$$
\times\left.\frac{x^2+4x\sigma+
\sigma^2}{(x-\sigma)^4}\right\}
+\frac{1}{6}b^3\left\{\biggl[2+
\frac{5}{2}
\kappa  
+\bar\kappa+3(1+C)
\left(1-\frac{\pi^2}{3}+\kappa\right) +(1+C)^3\biggr]\right.
$$
$$
\times \frac{x(x^2+4x+1)}{(x-1)^4}
+x\int\limits_{0}\limits^{1} d\sigma \,\biggl[
3\biggl(1-\frac{\pi^2}{3}+\kappa	 
+(1+C)^2\biggr)
\ln(-\ln\sigma)
$$
\begin{equation}
+3(1+C)\ln^2(-\ln\sigma)+\ln^3(-\ln\sigma)
\biggr]\frac{x^3+11x^2\sigma+11x\sigma^2+\sigma^3}{(x-\sigma)^5}
\Biggr\}.
\label{26}
\end{equation}
This formula gives the nonperturbative contributions in an explicit
form.

\section*{IV. BEHAVIOR OF THE NONPERTURBATIVE CONTRIBUTIONS AT 
$Q>\Lambda$}

Let us turn to the large $Q$ behavior of the  nonperturbative 
contributions. The following expansions appear to be useful 
($x>1\ge \sigma\ge 0$)
\begin{eqnarray}
\frac{1}{x-1}&=&\sum\limits_{n=1}\limits^{\infty}\frac{1}{x^n},
\,\,\,\,
\frac{x}{(x-1)^2}=\sum\limits_{n=1}\limits^{\infty}\frac{n}{x^n},
\,\,\,\,
\frac{x(x+\sigma)}{(x-\sigma)^3}=\sum\limits_{n=1}\limits^{\infty}
\frac{n^2\sigma^{n-1}}{x^n},
\nonumber\\
\frac{x(1+x)}{(x-1)^3}&=&\sum\limits_{n=1}\limits^{\infty}
\frac{n^2}{x^n}, \,\,\,\, 
\frac{x(x^2+4x\sigma+\sigma^2)}{(x-\sigma)^4}=\sum\limits_{n=1}
\limits^{\infty}\frac{n^3\sigma^{n-1}}{x^n},  \label{27}
\\
\frac{x(x^2+4x+1)}{(x-1)^4}&=&\sum\limits_{n=1}\limits^{\infty}
\frac{n^3}{x^n}, \,\,\,\, 
\frac{x(x^3+11x^2\sigma+11x\sigma^2+\sigma^3)}{(x-\sigma)^5}=
\sum\limits_{n=1}
\limits^{\infty}\frac{n^4\sigma^{n-1}}{x^n}.
\nonumber
\end{eqnarray}
Note that the coefficients in Eqs.~(\ref{27}) are monomials in powers
of $n$. Expanding Eq.~(\ref{26}) in the inverse powers of $x$ with
using Eqs.~(\ref{27})
we have
\begin{equation}
a_{an}^{npt}(x)=\sum\limits_{n=1}\limits^{\infty}\frac{c_n}{x^n},
\label{28}
\end{equation}
where
$$
c_n=-1+bn\left\{1+ C+n\int\limits_{0}\limits^{1}d\sigma\, 
\sigma^{n-1}
\ln\left(-\ln(\sigma)\right)\right\}
$$
$$
-\frac{1}{2}b^2n^2\left\{1+\kappa-\frac{\pi^2}{3}+(1+C)^2+n\int
\limits^1\limits_0 d\sigma\, \sigma^{n-1}\biggl[2(1+C)\ln\left(-\ln(
\sigma)\right)+\ln^2\left(-\ln(\sigma)\right)\biggr]\right\}
$$
$$
+\frac{1}{6}b^3n^3\left\{2+\frac{5}{2}\kappa  
+\bar\kappa+3(1+C)
\left(1-\frac{\pi^2}{3}+\kappa\right) +(1+C)^3
+n\int\limits_{0}\limits^{1} d\sigma \,\sigma^{n-1}\biggl[
3\biggl(1-\frac{\pi^2}{3}+\kappa	\right.
$$
\begin{equation}
+(1+C)^2\biggr)\ln(-\ln\sigma)
+3(1+C)\ln^2(-\ln\sigma)+\ln^3(-\ln\sigma)
\biggr]\Biggr\}.
\label{29}
\end{equation}
Making the change of variable $\sigma=\exp(-t)$ and integrating 
\cite{Bateman}, \cite{Prudnikov} over $t$ one can find
\begin{eqnarray}
\int\limits_{0}\limits^{1}d\sigma\, \sigma^{n-1}
\ln\left(-\ln(\sigma)\right)&=&\int\limits_{0}\limits^{\infty}dt\,
e^{-nt}\ln(t)=-\frac{1}{n}\left(\ln (n)+\gamma\right),
\nonumber\\
\int\limits_{0}\limits^{1}d\sigma\, \sigma^{n-1}
\ln^2\left(-\ln(\sigma)\right)&=&\int\limits_{0}\limits^{\infty}dt\,
e^{-nt}\ln^2(t)=\frac{1}{n}\left[\left(\ln (n)+\gamma\right)^2
+\frac{\pi^2}{6}\right],
\label{30}\\
\int\limits_{0}\limits^{1}d\sigma\, \sigma^{n-1}
\ln^3\left(-\ln(\sigma)\right)&=&\int\limits_{0}\limits^{\infty}dt\,
e^{-nt}\ln^3(t)=-\frac{1}{n}\left[\left(\ln (n)+\gamma\right)^3
+\frac{\pi^2}{2}\bigl(\ln(n)+\gamma\bigr)+2\zeta_3\right].
\nonumber
\end{eqnarray}
Here  $\gamma$ is the Euler constant,  $\gamma\simeq 0.5772$.
From Eqs.~(\ref{29}), (\ref{30}) we  finally  have
$$
c_n=-1+bn\left[1+C-\gamma-\ln(n)\right]-\frac{1}{2}b^2n^2\left[
1-\frac{\pi^2}{6}+\kappa+\Bigl(1+C-\gamma-\ln(n)\Bigr)^2\right]
$$
$$
+\frac{1}{6}b^3n^3\left[2+\frac{5}{2}\kappa+\bar\kappa-
2\zeta_3+\Bigl(1+C-\gamma-\ln (n)\Bigr)^3\right.
$$
\begin{equation}
+\left.3\Bigl(1+C-\gamma-\ln
(n)\Bigr)\left(1-\frac{\pi^2}{6}+\kappa\right)
\right].
\label{31}
\end{equation}
We can see from Eq.~(\ref{31})
that power series~(\ref{28}) is uniformly convergent at $x>1$
and its convergence radius is  equal to unity.
The resulting Eq.~(\ref{31}) is scheme independent in the sense
that $n_f$ dependence is not fixed here, and the method used above
allows one in principle to calculate next loops contributions to
clarify the general structure of the coefficients $c_n$.

For numerical evaluation of the coefficients $c_n$ we choose the
$\overline{MS}$ scheme values of $\kappa$, $\bar\kappa$ and
assume that $C=0$. Then the coefficients $c_n$ are dependent on $n$,
$n_f$, and on the number of loops
taken into account. In Table~\ref{t2} we give the values of $c_n$ and
loop corrections for $n_f=0,3,4,5,6$.
The 1-loop order contributions to $c_n$ are equal
to $-1$ for all $n$ and $n_f$. Up to 4-loop approximation the
coefficients $c_n$ for all  $n$, $n_f$ are negative. 
With the exception of the 3-loop case at $n_f=6$,
the 2 --- 4-loop
coefficients $c_n$ for $n_f=0,3,4,5,6$ monotonously increase in the
absolute value with increasing $n$.  
In the ultraviolet region ($x\gg 1$) the nonperturbative
contributions 
are determined by the first term of the series~(\ref{28}).
One can  see that for all $n_f$ up to four loops $c_1$ is 
of the order of unity.
The account for the higher loop corrections results in some 
compensation of the 1-loop  leading at large $x$ term of the
form $1/x$.

\begin{table}[tbp]
\caption{The dependence of $c_n$  and loop corrections on $n$ and
$n_f$ for the 1 --- 4-loop
cases.}
\label{t2}
\begin{center}
\begin{tabular}{r r r r r r r r r}\hline \hline
     &$n$& $c^{1-loop}_n$  &$\Delta_n^{2-loop}$  &
     $\Delta_n^{3-loop}$ 
	 & $\Delta_n^{4-loop}$& $c_n^{2-loop}$ &$c_n^{3-loop}$ &
	 $c_n^{4-loop}$\\ \hline
$n_f=0$& 1&-1.0
&   0.35640&  -0.01568&  -0.03900&  -0.64360&  -0.65929&  -0.69828\\
& 2&-1.0
&  -0.45582&   0.08741&  -0.16455&  -1.45582&  -1.36841&  -1.53296\\
& 3&-1.0
&  -1.70912&  -1.03012&  -0.89283&  -2.70912&  -3.73924&  -4.63207\\
& 4&-1.0
&  -3.24886&  -4.51236&  -5.11707&  -4.24886&  -8.76122& -13.87829\\
& 5&-1.0
&  -5.00160& -11.31238& -18.56032&  -6.00160& -17.31398& -35.87430\\
& 6&-1.0
&  -6.92407& -22.24971& -49.77660&  -7.92407& -30.17378& -79.95039\\
& 8&-1.0
& -11.17217& -59.34790&-213.31981& -12.17217& -71.52007&-284.83987\\
&10&-1.0
& -15.84626&-120.76945&-616.88776& -16.84626&-137.61571&-754.50348\\
	 \hline
$n_f=3$& 1&-1.0
&   0.33405&   0.01608&  -0.07825&  -0.66595&  -0.64987&  -0.72812\\
& 2&-1.0
&  -0.42724&   0.19624&  -0.37379&  -1.42724&  -1.23101&  -1.60480\\
& 3&-1.0
&  -1.60196&  -0.63626&  -1.28115&  -2.60196&  -3.23823&  -4.51937\\
& 4&-1.0
&  -3.04517&  -3.48651&  -5.07338&  -4.04517&  -7.53168& -12.60506\\
& 5&-1.0
&  -4.68801&  -9.19185& -16.30462&  -5.68801& -14.87987& -31.18449\\
& 6&-1.0
&  -6.48996& -18.47225& -41.82403&  -7.48996& -25.96221& -67.78624\\
& 8&-1.0
& -10.47171& -50.22832&-174.16411& -11.47171& -61.70003&-235.86414\\
&10&-1.0
& -14.85275&-103.11451&-499.79465& -15.85275&-118.96725&-618.76190\\
	 \hline
$n_f=4$& 1&-1.0
&   0.31252&   0.04949&  -0.11006&  -0.68748&  -0.63799&  -0.74805\\
& 2&-1.0
&  -0.39970&   0.31341&  -0.52880&  -1.39970&  -1.08630&  -1.61510\\
& 3&-1.0
&  -1.49872&  -0.23818&  -1.51417&  -2.49872&  -2.73690&  -4.25107\\
& 4&-1.0
&  -2.84891&  -2.48499&  -4.77483&  -3.84891&  -6.33389& -11.10872\\
& 5&-1.0
&  -4.38587&  -7.15989& -13.83276&  -5.38587& -12.54576& -26.37852\\
& 6&-1.0
&  -6.07168& -14.89306& -34.04904&  -7.07168& -21.96474& -56.01377\\
& 8&-1.0
&  -9.79681& -41.69613&-138.28745& -10.79681& -52.49294&-190.78040\\
&10&-1.0
& -13.89549& -86.71011&-394.96378& -14.89549&-101.60560&-496.56937\\
	 \hline
$n_f=5$& 1&-1.0
&   0.27813&   0.11653&  -0.16002&  -0.72187&  -0.60535&  -0.76537\\
& 2&-1.0
&  -0.35571&   0.55755&  -0.75021&  -1.35571&  -0.79817&  -1.54837\\
& 3&-1.0
&  -1.33377&   0.50736&  -1.78434&  -2.33377&  -1.82641&  -3.61075\\
& 4&-1.0
&  -2.53536&  -0.73077&  -4.12859&  -3.53536&  -4.26613&  -8.39472\\
& 5&-1.0
&  -3.90317&  -3.73728&  -9.82126&  -4.90317&  -8.64046& -18.46172\\
& 6&-1.0
&  -5.40344&  -9.01126& -22.11892&  -6.40344& -15.41470& -37.53362\\
& 8&-1.0
&  -8.71859& -28.07388& -85.51994&  -9.71859& -37.79247&-123.31240\\
&10&-1.0
& -12.36617& -60.94080&-243.69211& -13.36617& -74.30698&-317.99908\\
	 \hline
$n_f=6$& 1&-1.0
&   0.22433&   0.25378&  -0.22731&  -0.77567&  -0.52189&  -0.74920\\
& 2&-1.0
&  -0.28692&   1.07460&  -1.01673&  -1.28692&  -0.21231&  -1.22904\\
& 3&-1.0
&  -1.07581&   1.93179&  -2.00536&  -2.07581&  -0.14402&  -2.14938\\
& 4&-1.0
&  -2.04500&   2.37205&  -2.96183&  -3.04500&  -0.67295&  -3.63479\\
& 5&-1.0
&  -3.14826&   2.01775&  -4.07318&  -4.14826&  -2.13052&  -6.20370\\
& 6&-1.0
&  -4.35837&   0.54419&  -6.02095&  -5.35837&  -4.81418& -10.83512\\
& 8&-1.0
&  -7.03234&  -6.87467& -17.72700&  -8.03234& -14.90701& -32.63400\\
&10&-1.0
&  -9.97445& -21.85073& -53.77987& -10.97445& -32.82518& -86.60506\\
	 \hline \hline
\end{tabular}
\end{center}
\end{table}

\section*{V. MOMENTUM DEPENDENCE OF $\alpha_{an}$ }

The expansion  coefficients increase in the absolute value not too
fast and therefore the representation of the analytic running
coupling of QCD in the form
\begin{equation}
\alpha_{an}(Q^2)=\alpha^{pt}(Q^2)+\frac{4\pi}{b_0}\sum\limits^{\infty
}\limits_
{n=1}c_n\left(\frac{\Lambda^2}{Q^2}\right)^n,
\label{m14}
\end{equation}
with $c_n$ as in Eq.~(\ref{31}) 
provides one with the effective method for the calculation of 
$\alpha_{an}$ at $Q>\Lambda$.  At that there is no need for the
summation
of large number of terms of the series. 
Let us see what  the convergence properties of the
series~(\ref{28}) are. 
Since $pn>\ln^3(n)$ for all $n\ge 1$ and $p>p_0=(3/e)^3\simeq 1.4$,
one can consider the series
\begin{equation}
S_4=\sum\limits_{n=1}\limits^{\infty}\frac{n^4}{x^n}=
\frac{x(x^3+11x^2+11x+1)}{(x-1)^5}
\label{n1}
\end{equation}
as a comparison series for Eq.~(\ref{28}) with 
coefficients~(\ref{31}). The convergence properties of the series
Eq.~(\ref{28}) are not worse
than that for the series~(\ref{n1}). The absolute error for the
$N$-terms approximation of the series~(\ref{n1}) is
$$
\Delta_4^{(N)}=\frac{1}{x^N(x-1)}\left[\frac{x(x^3+11x^2+11x+1)}{(x-1
)^4}+\frac{4Nx(x^2+4x+1)}{(x-1)^3}\right.
$$
\begin{equation}
\left.+\frac{6N^2x(x+1)}{(x-1)^2}+\frac{4N^3x}{x-1} +N^4\right].
\label{n2}
\end{equation}
It is dependent on $Q$ and $N$ (with given
$n_f$, $\Lambda$) and  the larger $Q$ and $N$ are, the smaller it is. 
For rather small $x=2$ $(Q=1.4\Lambda)$  from Eq.~(\ref{n2}) we find
that the error of the approximation of the series~(\ref{n1}) for 
$N=50$ is $\simeq  10^{-9}$ and for $N=100$ it is
$\simeq 10^{-23}$. For larger $Q$ there are no reasons to sum a large
number of terms.
For the approximation of the analytic running
coupling
with only one first term of the series~(\ref{28}) for the
nonperturbative
contributions taken into account,
$$
\alpha_{an}(Q^2)\simeq\alpha^{pt}(Q^2)-\frac{4\pi}{b_0}
\left\{1-b\left(1-\gamma\right)+\frac{1}{2}b^2\left(
1-\frac{\pi^2}{6}+\kappa+\left(1-\gamma\right)^2\right)\right.
$$
\begin{equation}
\left.
-\frac{1}{6}b^3\left[2+\frac{2}{5}\kappa+\bar\kappa-2\zeta_3+
\left(1-\gamma\right)^3+3\left(1-\gamma\right)\left(1-\frac{\pi^2}{6}
+\kappa\right)\right]
\right\}
\frac{\Lambda^2}{Q^2},
\label{n3}
\end{equation}
the relative error was studied in Ref.~\cite{JofP} for the 1 ---
4-loop order cases. The approximation of the nonperturbative "tail"
by the leading term has been shown to give a one percent accuracy for
$\alpha_{an}$ already at $Q\sim 5\Lambda$.

In Fig.~4 the $x$ dependencies
of $a_{an}$, $a^{pt}$, $a_{an}^{npt}$  are presented
for the 1 --- 4-loop order cases. The nonperturbative contributions
have been calculated by the series summation
and the analytic running coupling has been calculated through
the dispersive representation~(\ref{2}). It turned out that
the numerical integration in the cases considered expects
definite caution. Insufficient accuracy of integration can
look as an ungrounded stability of the analytic running coupling
behavior
with increase of the order of approximation.
The equality (with the accuracy of
$2\cdot 10^{-3}$ percent) of  $a_{an}(x)$  calculated through the
dispersive representation	and the sum of 	 $a^{pt}(x)$ and
$a_{an}^{npt}$ calculated as the series for all $x$ from 2 to 20
served us as a criterion for the integration precision.
The perturbative component  $a^{pt}$ increases with the decrease of
$x$  
reaching unity at $x\sim 3$ ($Q\sim 1.7\Lambda$). The nonperturbative
component
is negative (at $x>1$), it decreases with $x$ compensating for
the increase of the perturbative component. According to 
representation~(\ref{2}) the quantity  $a_{an}(x)$ is regular
for all  $x>0$ and  $a^{(l)}_{an}(0)=1$ ($l$ is the number of loops
of the approximation).
Though the derivative of $a_{an}(x)$ is infinite at zero  we however
make sure numerically of the higher loop stability of $a_{an}(x)$ in
the
infrared
region.  As seen in Fig.~4, the 3-loop and 4-loop analytic curves
practically coincide even before the normalization at some finite
point.    As for the corresponding perturbative curves which have no
common point at zero, they are not close to each other already at
$x<5$.

Let us consider the momentum dependence of $\alpha_{an}$ (and
$\alpha^{pt}$ for comparison) in the low momentum region provided
that all solutions are normalized at the central point of
the world average
value $\alpha(M^2_Z)=0.1181$ $\pm 0.002$~\cite{Data}. In this case
the heavy quark thresholds should be taken into account. It seems
natural to demand the analytic running coupling  be continuous across
thresholds. Let us adopt for $\alpha_{an}$, $\alpha^{pt}$ and for all
1 --- 4-loop order cases the  
normalization condition
$\alpha^{(n_f=5)}(M_Z^2)=$ 0.1181, $M_Z=$ 91.1882 GeV and the
matching conditions
$\alpha^{(n_f=5)}(m_b^2)=$ $\alpha^{(n_f=4)}(m_b^2)$, $m_b=$ 4.3 GeV
and $\alpha^{(n_f=4)}(m_c^2)=$ $\alpha^{(n_f=3)}(m_c^2)$, $m_c=$ 1.3
GeV~\footnote{The matching conditions  sensitivity of $\alpha_{an}$
will be  considered slightly later.}.
The corresponding sets of parameters $\Lambda$ are given in
Table~3. 
As seen in Table~3, $\Lambda_{an}^{(n_f=5)}\simeq $
$\Lambda_{pt}^{(n_f=5)}$, since the nonperturbative
contributions in fact die out at the scale of normalization
\footnote{If to normalize the solutions as
$\alpha^{(n_f=4)}(M_\tau^2)=$ 0.35, $M_\tau=$ 1777.03
MeV~\cite{Data} with the same matching conditions one obtains
substantially larger
values of $\Lambda_{an}$  (e.g., for the 4-loop case
$\Lambda_{an}^{(n_f=3, 4,
5)}\simeq $ 630 MeV,
490 MeV, 350 MeV, respectively).  At that the
higher loop stability is also observed,
for the 2 --- 4-loop cases $\alpha_{an}^{(n_f=5)}(M_Z^2)=$
0.128.}.
The momentum dependence of $\alpha_{an}$, $\alpha^{pt}$  for 
the 1 --- 4-loop order cases 
is presented
in Fig.~5. As seen from  Fig.~5, the 2 --- 4-loop curves for
$\alpha_{an}$ practically coincide indicating the
higher loop stability of the analytic running coupling. 
The corresponding values of 
$\alpha_{an}^{(n_f=4)}(M_{\tau}^2)$, 
$\alpha_{pt}^{(n_f=4)}(M_{\tau}^2)$ for the 1 --- 4-loop cases
are given in Table~4.

%  We go from M_Z  c=   0.00000   nryada=100
%               nf=  3,4,5   
%made of an11.dat;18  
\begin{table}[tbp]
\caption{
%Series summation. Improved.
The parameters $\Lambda_{pt}^{(n_f)}$ (MeV), 
$\Lambda_{an}^{(n_f)}$ (MeV). $n_f$ is the number of active quark
flavors, the number of loops is indicated.
The normalization and matching conditions are 
$\alpha^{(n_f=5)}(M_Z^2)=$ 0.1181, $M_Z=$ 91.1882 GeV,
$\alpha^{(n_f=5)}(m_b^2)=$ $\alpha^{(n_f=4)}(m_b^2)$, $m_b=$ 4.3
GeV,
$\alpha^{(n_f=4)}(m_c^2)=$ $\alpha^{(n_f=3)}(m_c^2)$, $m_c=$ 1.3
GeV.}
\label{series}
\begin{center}
\begin{tabular}{c| c c c c}\hline \hline
     &1-loop &2-loop&3-loop&4-loop
	 \\ \hline
$\Lambda_{pt}^{(n_f=3)}$  
 &     143.77 &     372.50 &     328.98 &     332.50
\\ \hline
$\Lambda_{pt}^{(n_f=4)}$ 
 &     120.55 &     325.91 &     289.67 &     291.39
\\ \hline 
$\Lambda_{pt}^{(n_f=5)}$  
 &      88.35 &     227.51 &     209.54 &     209.53
%          Lambdaan(n_f,nloops)
\\ \hline \hline
$\Lambda_{an}^{(n_f=3)}$  
 &     150.64 &     454.21 &     382.30 &     389.50
\\ \hline
$\Lambda_{an}^{(n_f=4)}$  
 &     121.61 &     339.90 &     298.91 &     301.64
\\ \hline
$\Lambda_{an}^{(n_f=5)}$  
 &      88.35 &     227.60 &     209.60 &     209.61
\\  \hline \hline
\end{tabular}
\end{center}
\end{table}

%    Table,  an7.dat;3 is used
%    We go from M_Z  c=   0.00000   nryada= 30
%               nf=  3,4,5   
%made of an7.dat;3   No integration, we use the series

\begin{table}[tbp]
%Series summation $(N=30)$.
\caption{The values
 $\alpha_{an}^{(n_f=4)}(M_{\tau}^2)$, 
$\alpha_{pt}^{(n_f=4)}(M_{\tau}^2)$ with $M_{\tau}=1777.03$ MeV. 
The normalization and matching conditions are
$\alpha^{(n_f=5)}(M_Z^2)=$ 0.1181, $M_Z=$ 91.1882 GeV,
$\alpha^{(n_f=5)}(m_b^2)=$ $\alpha^{(n_f=4)}(m_b^2)$, $m_b=$ 4.3
GeV.}
\label{t12}
\begin{center}
\begin{tabular}{r| c c c c}\hline \hline
     & 1-loop & 2-loop& 3-loop& 4-loop
	 \\ \hline
 $\alpha_{an}$ & 0.2740 &       0.2930 &       0.2943 &
 0.2943\\
\hline 
$\alpha^{pt}$  & 0.2802 &       0.3262 &       0.3179 &
0.3230\\
\hline \hline
\end{tabular}
\end{center}
\end{table}
Note that in Table~4 the values of 3-loop and 4-loop 
$\alpha_{an}(M_{\tau}^2)$ are the same and the 4-loop
$\alpha_{pt}(M_{\tau}^2)$ coincides with  $\alpha_s(M_{\tau})$ of
Ref.~\cite{Bethke} from $\tau$ decays. Thus, the extrapolation 
to the energy scale $M_Z$ using the 4-loop solution for $\alpha_s$
with
3-loop matching at the bottom quark pole mass $M_b=$ 4.7 GeV 
made in Ref.~\cite{Bethke} 
results in $\alpha_s(M_Z)=$ 0.1181. This value is used in our
calculations.

Let us consider the threshold matching conditions  sensitivity of
$\alpha_{an}$. 
We fix $\alpha^{(n_f=5)}(M_Z^2)$ at its world average value and
vary
the value of the matching point $\mu_b$ corresponding to the
$b$-quark
threshold. Then from the matching condition
$\alpha^{(n_f=5)}(\mu_b^2)=$ $\alpha^{(n_f=4)}(\mu_b^2)$ we can find
the dependence of the parameters $\Lambda^{(n_f=4)}$ on the matching
point value.
For a rather wide interval of $\mu_b$ it is shown in Fig.~6.
We see that $\Lambda^{(n_f=4)}$ parameters for the analytic running
coupling
go
higher than that for the perturbative coupling. In the region of
4 -- 5 GeV the $\mu_b$-dependence of $\Lambda^{(n_f=4)}_{an}$ is
somewhat weaker than that of $\Lambda^{(n_f=4)}_{pt}$.
The dependencies of $\alpha_{an}(M_{\tau}^2)$,
$\alpha^{pt}(M_{\tau}^2)$
on the matching point $\mu_b$ for the 1 --- 4-loop order cases are
shown in Fig.~7. For the analytic
coupling the curves go lower than the corresponding curves for the
perturbative coupling. 
The analytic coupling is much more stable than the perturbative
one with respect to higher loop corrections.
In the region of 4 -- 5 GeV the $\mu_b$-dependence of
$\alpha_{an}(M_{\tau}^2)$ is considerably weaker than that of
$\alpha^{pt}(M_{\tau}^2)$. In particular, for $\alpha(M_Z^2)=0.1181$,
$m_b=4.3\pm 0.2$ GeV
\begin{equation}
\alpha_{an}(M_{\tau}^2)=0.2943^{+0.0004}_{-0.0003},
\,\,\,\alpha^{pt}(M_{\tau}^2)=0.3230^{+0.0008}_{-0.0008}.
\label{match1}
\end{equation}
We give here the results for the 4-loop $\alpha_{an}$  and
$\alpha^{pt}$.
%file an116.dat;6

Let us consider one more heavy quark
threshold matching condition.
In the framework of the perturbation theory there is the
prescription~\cite{Chet97} to connect the couplings 
with different $n_f$ according to which the coupling can be
discontinuous at the matching point  $\mu_h$. The idea of an
implementation of this nontrivial matching conditions  is to make the
results (e.g., the connection between $\alpha(M_{\tau}^2)$ and
$\alpha(M_Z^2)$) be not substantially dependent on the exact value of
the matching point~\cite{Rodrigo}. This conditions take the most
simple form for two cases. First, $\mu_h=m_h\equiv m_h(m_h)$ where 
$m_h(\mu)$ is the  running $\overline{MS}$ mass of the heavy quark
and second, $\mu_h=M_h$ with $M_h$ being  the heavy quark pole mass.
Choosing the first one according
to~\cite{Chet97} we have
\begin{equation}
\alpha^{(n_f-1)}(\mu_h^2)=\alpha^{(n_f)}(\mu_h^2)
\label{12loop}
\end{equation}
for the 1-loop and 2-loop cases,
\begin{equation}
\alpha^{(n_f-1)}(\mu_h^2)=\alpha^{(n_f)}(\mu_h^2)\left[1+c_2\left
(\alpha^{(n_f)}
(\mu_h^2)/\pi\right)^2\right]
\label{3loop}
\end{equation}
for the 3-loop case, and
\begin{equation}
\alpha^{(n_f-1)}(\mu_h^2)=\alpha^{(n_f)}(\mu_h^2)\left[1+c_2\left
(\alpha^{(n_f)}
(\mu_h^2)/\pi\right)^2+c_3\left(\alpha^{(n_f)}(\mu_h^2)/\pi\right)^3
\right]
\label{4loop}
\end{equation}
for the 4-loop case. Here
\begin{equation}
c_2=\frac{11}{72},\; \;
c_3=\frac{564731}{124416}-\frac{82043}{27648}\zeta_3-\frac{2633}{3110
4}(n_f-1).
\label{c1c2}
\end{equation}
The notations for the coefficients in Eqs.~(\ref{3loop}) ---
(\ref{c1c2}) correspond to Ref.~\cite{Chet97}. The coefficients $c_n$
of the present paper have its own definition. 
For this variant of matching conditions
the dependencies of the parameters $\Lambda^{(n_f=4)}$  on the
matching
point value are close to those shown in Fig.~6 and  we do not give
the corresponding figure (according to Eq.~(\ref{12loop}) both
matching methods give the same results for the 1, 2-loop order
cases). The dependencies of $\alpha_{an}(M_{\tau}^2)$,
$\alpha^{pt}(M_{\tau}^2)$
on the matching point $\mu_b$ for the 3, 4-loop order cases are
also close to the previous case of the continuous matching shown
in Fig.~7. 
For $\alpha(M_Z^2)=0.1181$,
$m_b=4.3\pm 0.2$ GeV
\begin{equation}
\alpha_{an}(M_{\tau}^2)=0.2947^{+0.0003}_{-0.0003},
\,\,\,\alpha^{pt}(M_{\tau}^2)=0.3235^{+0.0008}_{-0.0004}.
\label{match2}
\end{equation}
The results are given  for the 4-loop case.
%file an116.dat;6
It is seen from Eqs.~(\ref{match1}), (\ref{match2}) that the values
of $\alpha_{an}(M_{\tau}^2)$ for two methods of matching are very
close to each other (this is true also for
$\alpha^{pt}(M_{\tau}^2)$).

In Fig.~8  
$\alpha_{an}(Q^2)$, $\alpha^{pt}(Q^2)$ are shown for the 1 --- 4-loop
order cases with the normalization condition
$\alpha^{(n_f=5)}(M_Z^2)=$ 0.1181, $M_Z=$ 91.1882 GeV and
continuous matching at $m_b=$ 4.3 GeV, $m_c=$ 1.3 GeV.
Without going into the details, we give in this figure the data  from
Table~6 of Ref.~\cite{Bethke} for the world summary of measurements
of $\alpha_s$.

%~\cite{Bethke} matches at pole masses $M_c=1.5$ GeV, $M_b=4.7$ GeV.
 
\section*{VI. CONCLUSIONS}

In contrast to the recent papers \cite{Magr} we apply the analytic
approach to the perturbative QCD running coupling constant in the
form of the standard expansion in the inverse powers of logarithms up
to
the four loop order. An introduction of the complex variable $t$ in
the spectral representation for the analytic running coupling and
study of the singularities structure of the integrand allowed one to
divide the analytic running coupling into perturbative (initial)
component and nonperturbative one (appeared as a consequence of
"forced" analyticity) exactly, as it is illustrated in Fig.~3.
These components turned out to be connected with different
singularities in the complex $t$-plane. It is shown that the
nonperturbative contributions can be represented in the form of the
expansion Eq.~(\ref{28}) in inverse powers of the momentum squared 
where the coefficients $c_n$ are defined by Eq.~(\ref{31}).
Eq.~(\ref{m14}) gives the effective method non-connected with
numerical integration for calculation of the analytic running
coupling  at $Q>\Lambda$ with the  calculation accuracy of
standard mathematical functions. It can be important 
for making popular the considered variant of $\alpha_{an}$.
In practice, for $Q$ corresponding
to $n_f=5$, it is sufficient to take account of the leading
nonperturbative term, as in Eq.~(\ref{n3}).

On the basis of the developed method we study the momentum dependence
of $\alpha_{an}$ giving at the same time the behavior of the
perturbative running coupling. To fix the solutions  we used for all
of them the same normalization condition at $M_Z$ where the
nonperturbative contributions are negligible quantities.
We can see in Fig.~4, Fig.~5, Fig.~7, and Fig.~8 the higher loop
stability of the
analytic running coupling for all $Q>0$ (the 1-loop case falls out
of the common picture). 
For the perturbative
case the higher loop stability takes place only at sufficiently large
$Q$.

We considered two variants of heavy quark threshold matching
conditions for $\alpha_{an}$.  
The results appeared to be very similar.
We showed the $b$-quark threshold matching conditions stability
of the analytic running coupling
\footnote{The $c$-quark threshold matching was considered in
Ref.~\cite{YadFiz} with same result.}.
As a criterion we  considered the dependence on the matching
point $\mu_b$ of the correspondence of $\alpha_{an}(M_{\tau}^2)$ 
to $\alpha_{an}(M_Z^2)$  for the 1 --- 4-loop
cases (for comparison $\alpha^{pt}$ was considered simultaneously).
The situation 
is illustrated by Fig.~7.
The energy scale evolution of the
analytic running coupling gives  $\alpha(M_\tau^2)=$
0.2943$^{+0.0004}_{-0.0003}$ for the normalization at the world
average
value of $\alpha(M_Z^2)=$ 0.1181
and matching by continuity with $m_b=$ 4.3 $\pm$ 0.2 GeV. With the
same normalization condition
at the scale of $M_Z$ for both matching methods considered 
$\alpha_{an}(M_{\tau}^2)$ is about 0.03 less than
$\alpha^{pt}(M_{\tau}^2)$. Therefore, if one regards $\alpha_{an}$ as
a true running coupling constant the noticeable discrepancy 
with $\tau$ lepton decay data~\cite{Data,Bethke} arises.

A possible solution of this problem can be found if one changes
the normalization condition at $M_Z$. Let it corresponds to the value
of Refs.~\cite{hepph0107282,hep0109084} $\alpha(M_Z^2)\simeq$
0.124 which 
is appreciably larger then the conventional one. Then for the 4-loop
case the result is
$\alpha_{an}(M_{\tau}^2)=$ $0.3270^{+0.0004}_{-0.0003}$ with
continuous matching at $m_b=4.3\pm 0.2$ GeV. 

As seen  in Fig.~8, the analytic approach gives the running
coupling which does not deviate essentially at sufficiently large
momentum values from the usual perturbative running coupling
constant. In the infrared region this approach allows one to solve
the principal difficulty connected with nonphysical singularities.
The question arises whether the approach described takes into account
the nonperturbative contributions to the right degree.
There is a whole series of the approaches in which the
nonperturbative  contributions to the Green functions and running
coupling are studied. These approaches are beyond the scope of the
present paper and we only point out  some
papers~\cite{AAA,Simonov,Zakharov,Grunberg,Nester,Sidor} dealing with
the
nonperturbative contributions
to the running coupling.

To summarize, the analytic running coupling seems to be a good basis
for the
problem of "genuine nonperturbative" contributions in "physical"
$\alpha_s$, which
needs further analysis.

\section*{ACKNOWLEDGMENTS}

I am deeply indebted to B.A.~Arbuzov, V.A.~Petrov, V.E.~Rochev,
D.V.~Shirkov for useful discussions. This work has been partly
supported by RFBR under Grant No.~02-01-00601.

\section*{APPENDIX}

We give here the identities we need in our computations which
can be obtained by means of an integration by parts. Let function
$f(z)$ of complex variable $z$ be regular in some domain $D$
where $z=0 \in D$. Dealing with the singularities of 
the integrands at the origin of the pole type coinciding
with the logarithmic type branch points we cut the domain $D$
along real negative semiaxis. Then for any contour $\tilde C$ in
the cut domain $\tilde D$ which goes from $z_1\neq 0$ to 
$z_2 \neq 0$  one can find
$$
\int\limits_{\tilde C} \frac{dz}{z} f(z)=-\int\limits_{\tilde C}
dz\,\ln(z) f'(z)+
\ln(z)f(z)\Big\vert^{z_2}_{z_1},
$$
$$
\int\limits_{\tilde C} \frac{dz}{z^2} f(z)=-\int\limits_{\tilde C}
dz\,\ln(z)
f''(z)+
\left\{-\frac{1}{z}f(z)+\ln(z)f'(z)\right\}\Big\vert^{z_2}_{z_1},
$$
$$
\int\limits_{\tilde C} \frac{dz}{z^3}
f(z)=-\frac{1}{2}\int\limits_{\tilde C} dz\,
\ln(z) f'''(z)+
\left\{-\frac{1}{2z^2}f(z)-\frac{1}{2z}f'(z)+\frac{1}{2}\ln(z)f''(z)
\right\}\Big\vert^{z_2}_{z_1},
$$
$$
\int\limits_{\tilde C} \frac{dz}{z^4}
f(z)=-\frac{1}{6}\int\limits_{\tilde C} dz\,
\ln(z) f''''(z)+
\left\{-\frac{1}{3z^3}f(z)-\frac{1}{6z^2}f'(z)-\frac{1}{6z}f''(z)
+\frac{1}{6}\ln(z)f'''(z)
\right\}\Big\vert^{z_2}_{z_1},
$$
$$
\int\limits_{\tilde C} \frac{dz}{z}\ln(z)
f(z)=-\frac{1}{2}\int\limits_{\tilde C} 
dz\,\ln^2(z) f'(z)+
\frac{1}{2}\ln^2(z)f(z)\Big\vert^{z_2}_{z_1},
$$
$$
\int\limits_{\tilde C} \frac{dz}{z^2}\ln(z) f(z)=-\int\limits_{\tilde
C} dz\,\left(
\ln(z)+\frac{1}{2}\ln^2(z)\right) f''(z)+
\left\{-\frac{1}{z}f(z)-\frac{\ln(z)}{z}f(z)+\ln(z)f'(z)\right.
$$
$$
+\left.\frac{1}{2}
\ln^2(z)f'(z)\right\}\Big\vert^{z_2}_{z_1},
$$
$$
\int\limits_{\tilde C} \frac{dz}{z^3}\ln(z) f(z)=-\int\limits_{\tilde
C} dz\,\left(
\frac{3}{4}\ln(z)+\frac{1}{4}\ln^2(z)\right) f'''(z)+
\left\{-\frac{1}{4z^2}f(z)-\frac{3}{4z}f'(z)-\frac{\ln(z)}{2z^2}f(z)
\right.
$$
$$
-\left.\frac{\ln(z)}{2z}f'(z)+\frac{3\ln(z)}{4}f''(z)+\frac{\ln^2(z)}
{4}f''(z)\right\}\Big\vert^{z_2}_{z_1},
$$
$$
\int\limits_{\tilde C} \frac{dz}{z^4}\ln(z) f(z)=-\int\limits_{\tilde
C} dz\,\left(
\frac{11}{36}\ln(z)+\frac{1}{12}\ln^2(z)\right) f''''(z)+
\left\{-\frac{1}{9z^3}f(z)-\frac{5}{36z^2}f'(z)
\right.
$$
$$
-\frac{11}{36z}f''(z)-\left.\frac{\ln(z)}{3z^3}f(z)-\frac{\ln(z)}{6z^
2}f'(z)-\frac
{\ln(z)}{6z}
f''(z)+\frac{11\ln(z)}{36}f'''(z)+\frac{\ln^2(z)}{12}f'''(z)
\right\}\Big\vert^{z_2}_{z_1},
$$
$$
\int\limits_{\tilde C} \frac{dz}{z}\ln^2(z)
f(z)=-\frac{1}{3}\int\limits_{\tilde C}
dz\,\ln^3(z) f'(z)+
\frac{1}{3}\ln^3(z)f(z)\Big\vert^{z_2}_{z_1},
$$
$$
\int\limits_{\tilde C} \frac{dz}{z^2}\ln^2(z)
f(z)=-\int\limits_{\tilde C} dz\,
\left(
2\ln(z)+\ln^2(z)+\frac{1}{3}\ln^3(z)\right) f''(z)+
\left\{-\frac{2}{z}f(z)-\frac{2\ln(z)}{z}f(z)
\right.
$$
$$
+\left.2\ln(z)f'(z)-\frac{\ln^2(z)}{z}f(z)+\ln^2(z)f'(z)+\frac
{\ln^3(z)}
{3}f'(z)\right\}\Big\vert^{z_2}_{z_1},
$$
$$
\int\limits_{\tilde C} \frac{dz}{z^3}\ln^2(z)
f(z)=-\int\limits_{\tilde C} dz\,
\left(
\frac{7}{4}\ln(z)+\frac{3}{4}\ln^2(z)+\frac{1}{6}\ln^3(z)\right) 
f'''(z)+
\left\{-\frac{1}{4z^2}f(z)\right. 
$$
$$
-\frac{7}{4z}f'(z)-\frac{\ln(z)}{2z^2}f(z)
-\frac{3\ln(z)}{2z}f'(z)+\frac{7\ln(z)}{4}f''(z)
$$
$$
-\left. \frac{\ln^2(z)}
{2z^2}f(z)-\frac{\ln^2(z)}
{2z}f'(z)+\frac{3\ln^2(z)}
{4}f''(z)+\frac{\ln^3(z)}
{6}f''(z)\right\}\Big\vert^{z_2}_{z_1},
$$
$$
\int\limits_{\tilde C} \frac{dz}{z^4}\ln^2(z)
f(z)=-\int\limits_{\tilde C} dz\,
\left(\frac{85}{108}\ln(z)+\frac{11}{36}\ln^2(z)+\frac{1}{18}\ln^3(z)
\right) f''''(z)+\left\{-\frac{2}{27z^3}f(z)\right. 
$$
$$
-\frac{19}{108z^2}f'(z)-\frac{85}{108z}f''(z)-\frac{2\ln(z)}{9z^3}
f(z)-\frac{5\ln(z)}{18z^2}f'(z)-\frac{11\ln(z)}{18z}f''(z)+
\frac{85\ln(z)}{108}f'''(z)
$$
$$
-\left. \frac{\ln^2(z)}{3z^3}f(z)-\frac{\ln^2(z)}{6z^2}f'(z)-\frac
{\ln^2(z)}{6z}f''(z)+\frac{11\ln^2(z)}
{36}f'''(z)+\frac{\ln^3(z)}{18}f'''(z)\right\}\Big\vert^{z_2}_{z_1},
$$
$$
\int\limits_{\tilde C} \frac{dz}{z}\ln^3(z)
f(z)=-\frac{1}{4}\int\limits_{\tilde C} 
dz\,\ln^4(z) f'(z)+
\frac{1}{4}\ln^4(z)f(z)\Big\vert^{z_2}_{z_1},
$$
$$
\int\limits_{\tilde C} \frac{dz}{z^2}\ln^3(z)
f(z)=-\int\limits_{\tilde C} dz\,
\left(6\ln(z)+3\ln^2(z)+\ln^3(z)
+\frac{1}{4}\ln^4(z)\right) f''(z)
$$
$$
+\left\{-\frac{6}{z}f(z) 
-\frac{6\ln(z)}{z}f(z)+6\ln(z)
f'(z)-\frac{3\ln^2(z)}{z}f(z)	  \right.
$$
$$
+\left.3\ln^2(z)f'(z)-\frac{\ln^3(z)}{z}f(z)+
\ln^3(z)f'(z)+\frac{\ln^4(z)}{4}f'(z)\right\}\Big\vert^{z_2}_{z_1},
$$
$$
\int\limits_{\tilde C} \frac{dz}{z^3}\ln^3(z)
f(z)=-\int\limits_{\tilde C} dz\,
\left(\frac{45}{8}\ln(z)+\frac{21}{8}\ln^2(z)+\frac{3}{4}\ln^3(z)
+\frac{1}{8}\ln^4(z)\right) f'''(z)
$$
$$
+\left\{-\frac{3}{8z^2}f(z) 
-\frac{45}{8z}f'(z)-\frac{3\ln(z)}{4z^2}f(z)-\frac{21\ln(z)}{4z}
f'(z)+\frac{45\ln(z)}{8}f''(z)-\frac{3\ln^2(z)}{4z^2}f(z)	  \right.
$$
$$
-\left.\frac{9\ln^2(z)}{4z}f'(z)+ \frac{21\ln^2(z)}{8}f''(z)-\frac
{\ln^3(z)}{2z^2}f(z)-\frac{\ln^3(z)}{2z}f'(z)+\frac{3\ln^3(z)}
{4}f''(z)+\frac{\ln^4(z)}{8}f''(z)\right\}\Big\vert^{z_2}_{z_1},
$$
$$
\int\limits_{\tilde C} \frac{dz}{z^4}\ln^3(z)
f(z)=-\int\limits_{\tilde C} dz\,
\left(\frac{575}{216}\ln(z)+\frac{85}{72}\ln^2(z)+\frac{11}{36}
\ln^3(z)+\frac{1}{24}\ln^4(z)\right) f''''(z)
$$
$$
+\left\{-\frac{2}{27z^3}f(z) 
-\frac{65}{216z^2}f'(z)-\frac{575}{216z}f''(z)-\frac{2\ln(z)}{9z^3}
f(z)-\frac{19\ln(z)}{36z^2}
f'(z)-\frac{85\ln(z)}{36z}f''(z)\right.
$$
$$
+\frac{575\ln(z)}{216}f'''(z)-\frac{\ln^2(z)}{3z^3}f(z)- 
\frac{5\ln^2(z)}{12z^2}f'(z)
-\frac{11\ln^2(z)}{12z}f''(z)  +\frac{85\ln^2(z)}{72}f'''(z)
-\frac
{\ln^3(z)}{3z^3}f(z)
$$
$$
-\left.\frac{\ln^3(z)}{6z^2}f'(z)-\frac{\ln^3(z)}
{6z}f''(z)+\frac{11\ln^3(z)}{36}f'''(z)+\frac{\ln^4(z)}{24}f'''(z)
\right\}\Big\vert^{z_2}_{z_1}.
$$

%
%                 Fig. 1
\begin{figure}[p]
\centerline{\psfig{figure=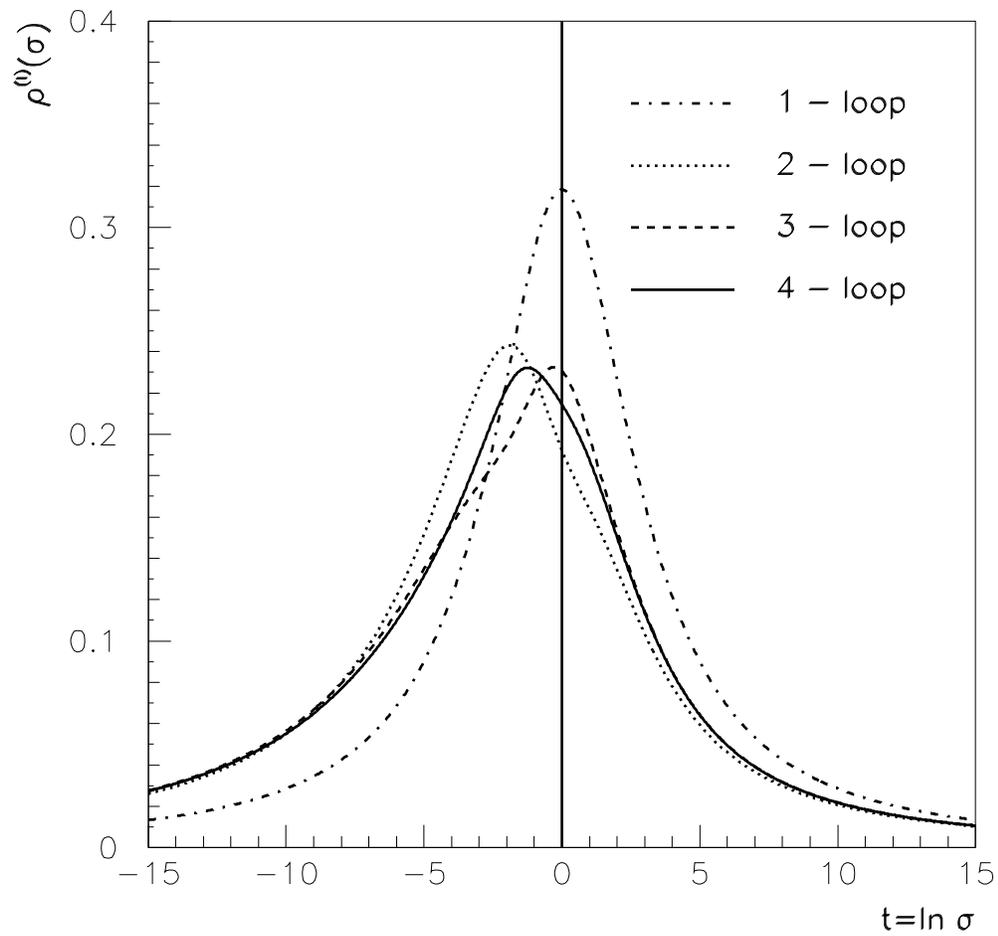,height=15cm,width=15cm}}
\caption{The spectral density of the analytic running coupling up to
four loop order.}
\label{fig2}
\end{figure}
%
%                 Fig. 2
\begin{figure}[p]
\centerline{\psfig{figure=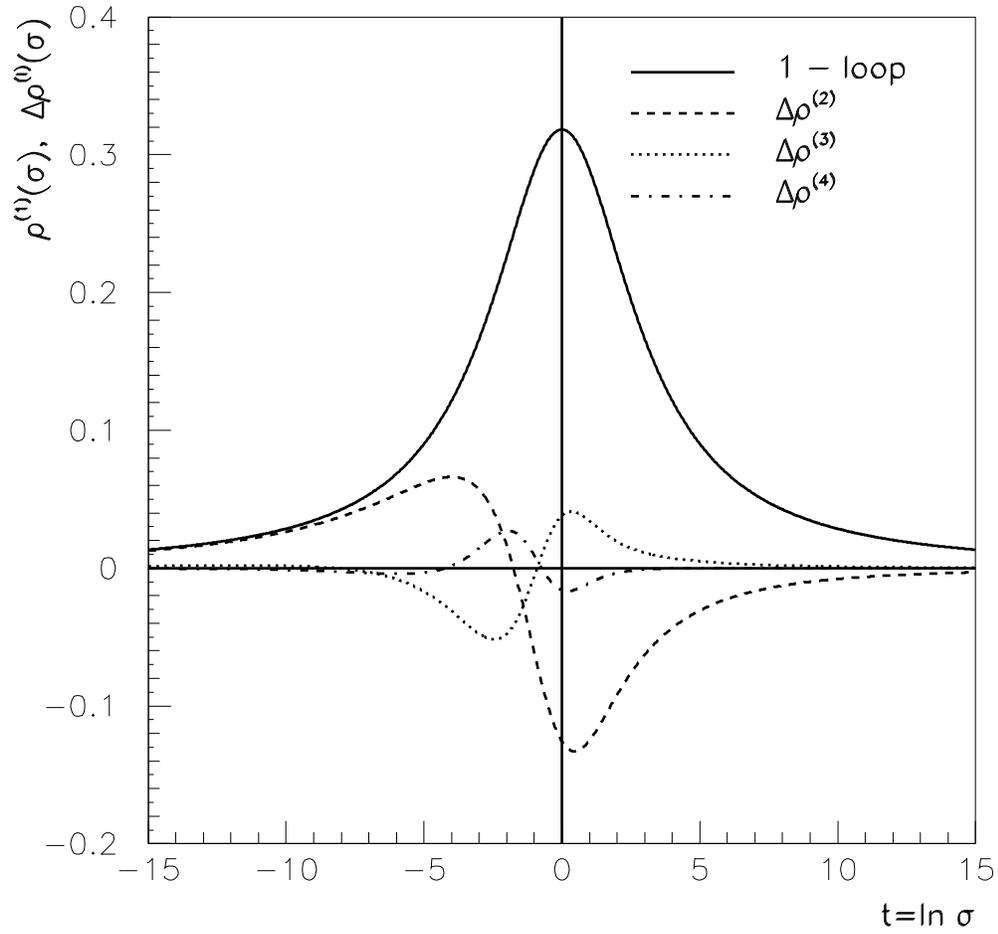,height=15cm,width=15cm}}
\caption{The higher loop order corrections for the spectral density.}
\label{fig1}
\end{figure}
%
%                 Fig. 3
\begin{figure}[p]
\centerline{\psfig{figure=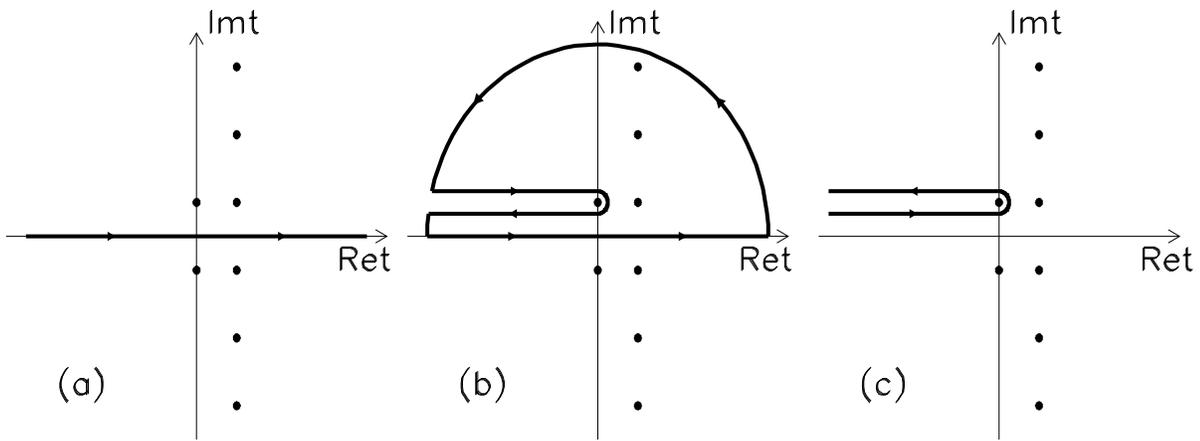,height=10cm,width=20cm}}
\caption{Complex t-plane integration. Perturbative contributions
arise from the poles at $t=\ln x+i\pi(1+2n)$, $n=0,1,2,...\,$.
Nonperturbative contributions emerge from the singularities at
$t=i\pi$.}
\label{fig3}
\end{figure}
%
%                 Fig. 4
\begin{figure}[p]
\centerline{\psfig{figure=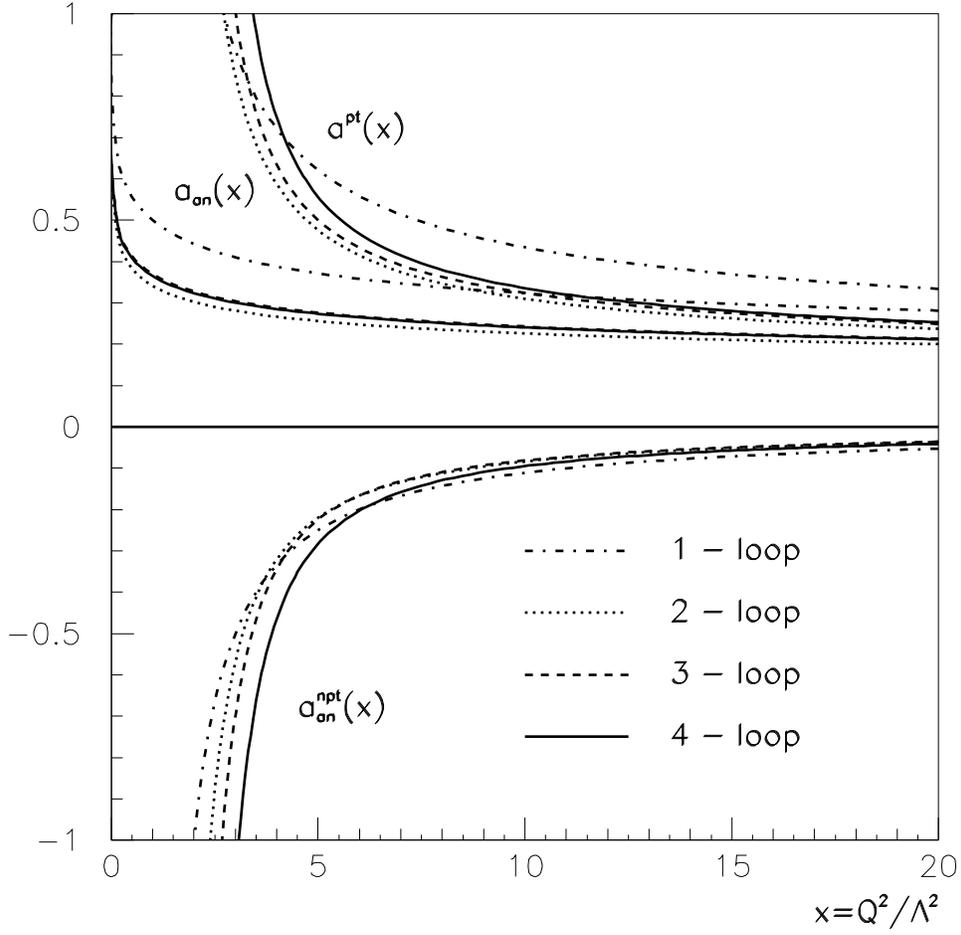,height=15cm,width=15cm}}
\caption{ The analytic running coupling $a_{an}$ and its
perturbative
component $a^{pt}$ and nonperturbative component $a_{an}^{npt}$ as
functions
of $x=Q^2/\Lambda^2$ for the 1 --- 4-loop order cases. Here $n_f=3$.}
\label{fig4}
\end{figure}
%
%
%
%
%                 Fig. 8
\begin{figure}[p]
\centerline{\psfig{figure=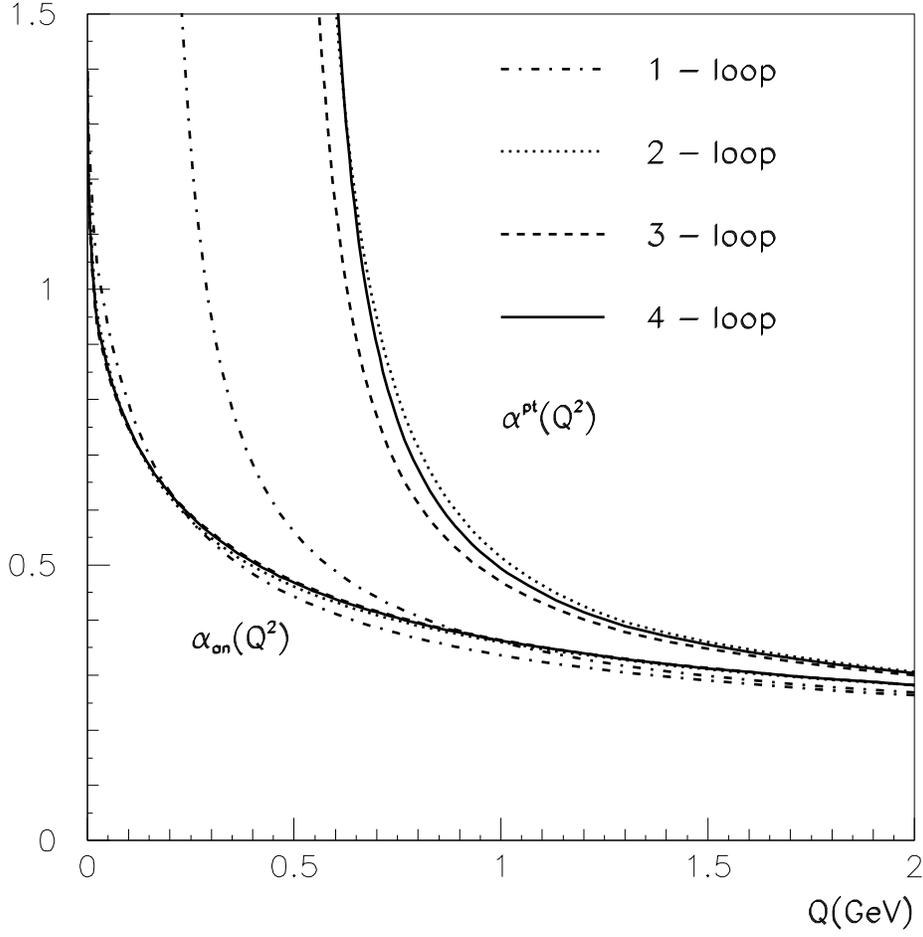,height=15cm,width=15cm}}
\caption{The momentum dependence of $\alpha_{an}$, $\alpha^{pt}$  for 
the 1 --- 4-loop order cases. 
The normalization and matching conditions are
$\alpha^{(n_f=5)}(M_Z^2)=$ 0.1181, $M_Z=$ 91.1882 GeV;
$\alpha^{(n_f=5)}(m_b^2)=$ $\alpha^{(n_f=4)}(m_b^2)$, $m_b=$ 4.3 GeV; 
$\alpha^{(n_f=4)}(m_c^2)=$ $\alpha^{(n_f=3)}(m_c^2)$, $m_c=$ 1.3
GeV.}
\label{fig8}
\end{figure}
%
%                 Fig. 9
\begin{figure}[p]
\centerline{\psfig{figure=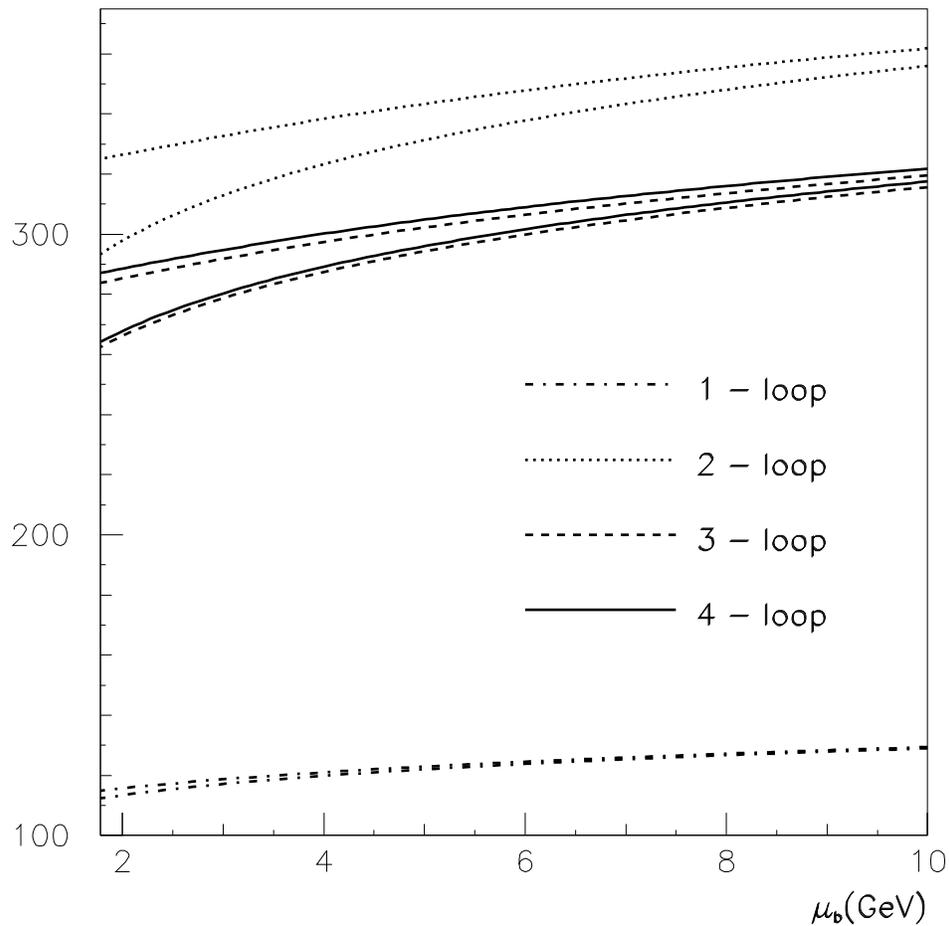,height=15cm,width=15cm}}
\caption{The dependencies of $\Lambda_{an}^{(n_f=4)}$,
$\Lambda_{pt}^{(n_f=4)}$ on the
matching point $\mu_b$ for the 1 --- 4-loop order cases. For the
analytic
coupling the curves go above the corresponding curves for the
perturbative coupling.
The normalization and matching conditions are
$\alpha^{(n_f=5)}(M_Z^2)=$ 0.1181, $M_Z=$ 91.1882 GeV;
$\alpha^{(n_f=5)}(\mu_b^2)=$ $\alpha^{(n_f=4)}(\mu_b^2)$.}
\label{fig9}
\end{figure}
%
%                 Fig. 10
\begin{figure}[p]
\centerline{\psfig{figure=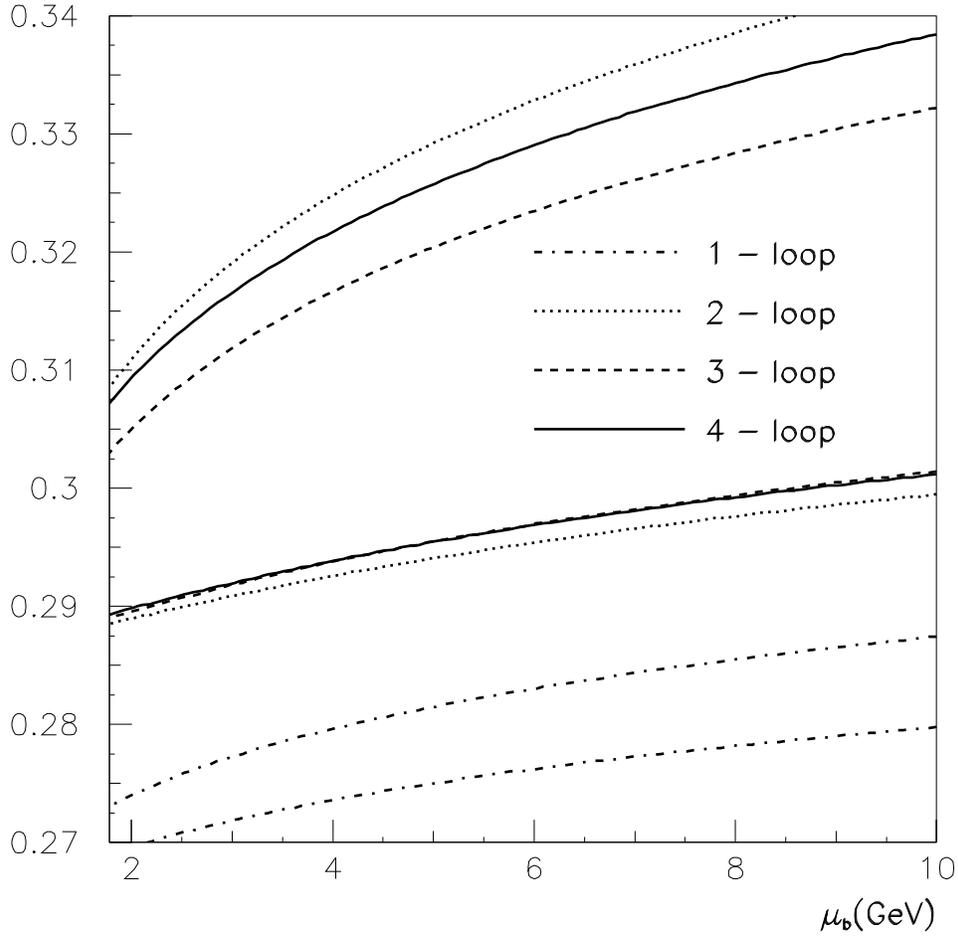,height=15cm,width=15cm}}
\caption{The dependencies of $\alpha_{an}(M_{\tau}^2)$,
$\alpha^{pt}(M_{\tau}^2)$
on the matching point $\mu_b$ for the 1 --- 4-loop order cases. For
the analytic coupling the curves go lower than the corresponding
curves for the perturbative coupling. The normalization and matching
conditions are
$\alpha^{(n_f=5)}(M_Z^2)=$ 0.1181, $M_Z=$ 91.1882 GeV;
$\alpha^{(n_f=5)}(\mu_b^2)=$ $\alpha^{(n_f=4)}(\mu_b^2)$.}
\label{fig10}
\end{figure}
%
%
%                 Fig. 13
\begin{figure}[p]
\centerline{\psfig{figure=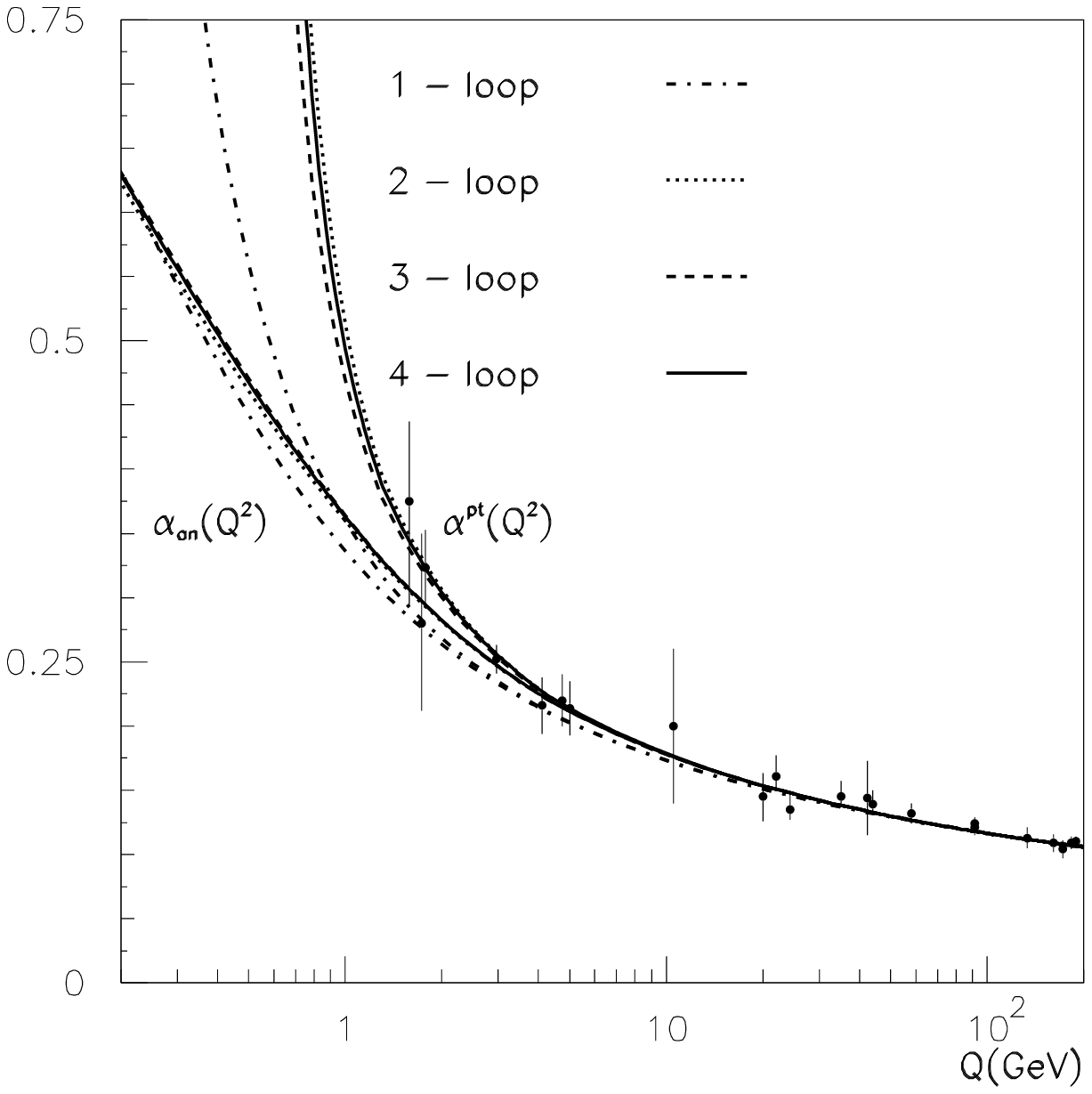,height=15cm,width=15cm}}
\caption{The analytic and perturbative couplings $\alpha_{an}(Q^2)$,
$\alpha^{pt}(Q^2)$ for the 1 --- 4-loop
order cases.
The normalization conditions are
$\alpha^{(n_f=5)}(M_Z^2)=$ 0.1181, $M_Z=$ 91.1882 GeV;
$\alpha^{(n_f=5)}(m_b^2)=$ $\alpha^{(n_f=4)}(m_b^2)$, $m_b=$ 4.3 GeV; 
$\alpha^{(n_f=4)}(m_c^2)=$ $\alpha^{(n_f=3)}(m_c^2)$, $m_c=$ 1.3
GeV.}
\label{fig13}
\end{figure}

\begin{thebibliography}{99}
\bibitem{Data}
Particle Data Group, D.E.~Groom {\it et al.}, Eur. Phys. J. 
C {\bf 15}, 85
(2000).
\bibitem{Bethke}
S.~Bethke, J. Phys. G {\bf 26}, R27 (2000),
 hep-ex/0004021.
\bibitem{Biebel}
O.~Biebel, Phys. Rep. {\bf 340}, 165 (2001).
\bibitem{SolShirTMF}
I.L.~Solovtsov and D.V.~Shirkov, Teor. Mat. Fiz. {\bf 120},
482 (1999) 
[Theor. Math. Phys. {\bf 120}, 1210 (1999)].
\bibitem{Red}
P.J.~Redmond,  Phys. Rev.  {\bf 112}, 1404 (1958).
\bibitem{Bog}
N.N.~Bogolubov, A.A.~Logunov, and  D.V.~Shirkov, 
Zh. \'{E}ksp. Teor.Fiz. {\bf 37}, 805 (1959) 
[Sov. Phys. JETP
{\bf 10}, 574 (1960)].
\bibitem{Shir}
D.V.~Shirkov and  I.L.~Solovtsov,   JINR Rapid Comm. 
{\bf 2[76]-96}, 5 (1996).
\bibitem{Shir1}
D.V.~Shirkov and  I.L.~Solovtsov,  Phys. Rev. Lett. {\bf 79},
1209 (1997).
\bibitem{Doksh}
Yu.L.~Dokshitzer, G.~Marchesini, and  B.R.~Webber,
Nucl. Phys.  {\bf B469}, 93 (1996).
\bibitem{A}
A.I.~Alekseev,  Phys. Rev. D {\bf 61}, 114005 (2000),
hep-ph/9906304.
\bibitem{IHEP40}
A.I.~Alekseev, 
in {\it Proceedings of the XVth Workshop 
on High Energy Physics and Quantum Field Theory},
Tver, Russia, 2000, edited by M.N.~Dubinin and V.I.~Savrin 
(INP Moscow State University, Moscow, 2001),
hep-ph/0011178.
\bibitem{YadFiz}
A.I.~Alekseev,  
Yad. Fiz. {\bf 65}, 1722  (2002) [Phys. At. Nucl.
{\bf 65},1678 (2002)].
\bibitem{JofP}
A.I.~Alekseev,  J.  Phys. G {\bf 27}, L117 (2001),
hep-ph/0105338.
\bibitem{Gross}
D.J.~Gross and F.~Wilczek, Phys. Rev. Lett. {\bf 30}, 1343 (1973);
H.D.~Politzer, Phys. Rev. Lett. {\bf 30}, 1346 (1973).
\bibitem{2-loop}
D.R.T.~Jones, Nucl. Phys. {\bf B75}, 531 (1974);
W.E.~Caswell, Phys. Rev. Lett. {\bf 33}, 244 (1974);
E.Sh.~Egoryan and O.V.~Tarasov, Teor. Mat. Fiz. {\bf 41}, 26 (1979) 
[Theor. Math. Phys. {\bf 41}, 863 (1979)].
\bibitem{3-loop1}
O.V.~Tarasov, A.A.~Vladimirov, and A.Yu.~Zharkov, Phys. Lett. 
B {\bf 93}, 429 (1980);
S.A.~Larin and J.A.M.~Vermaseren, Phys. Lett. B {\bf 303},
334 (1993).
\bibitem{Larin97}
T.~van~Ritbergen, J.A.M.~Vermaseren, and S.A.~Larin, Phys. Lett.
B {\bf 400}, 379 (1997);
J.~Ellis, M.~Karliner, and M.A.~Samuel, Phys. Lett. B {\bf 400}, 
176 (1997).
\bibitem{Bardeen78}
W.A.~Bardeen, A.~Buras, D.~Duke, and T.~Muta, Phys. Rev. D {\bf 18},
3998 (1978).
\bibitem{Chet97}
K.G.~Chetyrkin, B.A.~Kniehl, and M.~Steinhauser, Rhys. Rev. Lett.
{\bf 79}, 2184 (1997).
\bibitem{Marciano}
W.J.~Marciano, Phys. Rev. D {\bf 29}, 580 (1984).
\bibitem{Bateman}
H.~Bateman and A.~Erdelyi, {\it Higher transcendental functions}
Vol. 1\\
(McGraw-Hill, New York, 1953).
\bibitem{Prudnikov}
A.P.~Prudnikov, Yu.A.~Brychkov, and O.I.~Marichev, {\it Integrals
and Series} Vol. 1 \\
(Gordon and Breach, New York, NY, 1986).
\bibitem{Rodrigo}
G.~Rodrigo and A.~Santamaria, Phys. Lett. B {\bf 313}, 441
(1993).
\bibitem{Magr}
B.A.~Magradze,  Int. J.  Mod. Phys. A {\bf 15}, 2715  (2000),
hep-ph/9911456; B.A.~Magradze, hep-ph/0010070; D.S.~Kourashev and
B.A.~Magradze, hep-ph/0104142.
\bibitem{hepph0107282}
D.V.~Shirkov,  Eur. Phys. J. C {\bf 22}, 331
(2001).
%, hep-ph/0107282.
\bibitem{hep0109084}
J.H.~K\"{u}hn and M.~Steinhauser, Nucl. Phys. {\bf B619}, 588 (2001).
%DESY 01-130, hep-ph/0109084.
\bibitem{AAA}
A.I.~Alekseev and B.A.~Arbuzov,  Yad. Fiz. {\bf 61}, 314 (1998)
[Phys. At. Nucl.
{\bf 61},264. (1998)];
A.I.~Alekseev and B.A.~Arbuzov,  Mod. Phys. Lett. A {\bf 13}, 1447
(1998); A.I.~Alekseev, in {\it
Proceedings  of the  Workshop on   Nonperturbative	 Methods in
Quantum 
Field Theory},  Adelaide, Australia, 1998, edited by A.W.~Schreiber, 
A.G.~Williams, and  A.W.~Thomas  (World Scientific, Singapore, 1998),
hep-ph/9808206.
\bibitem{Simonov}
Yu.A.~Simonov, Pis'ma Zh.  \'{E}ksp. Teor.Fiz. {\bf 57}, 513
(1993) 
[JETP Lett.
{\bf 57}, 525 (1993)];
Yad. Fiz. {\bf 58}, 113  (1995) [Phys. At. Nucl.
{\bf 58},107 (1995]; Yad. Fiz. {\bf 65}, 140  (2002).
\bibitem{Zakharov}
R.~Akhoury and V.I.~Zakharov, hep-ph/9705318; V.I.~Zakharov, Prog.
Theor. Phys. Suppl. {\bf 131}, 107 (1998); Nucl. Phys. Proc. Suppl.
{\bf 74}, 392 (1999).
\bibitem{Grunberg}
G.~Grunberg, hep-ph/9705460, hep-ph/9807494;
K.A.~Milton, I.L.~Solovtsov, O.P.~Solovtsova, and V.I.~Yasnov, Eur.
Phys. J. C {\bf 14}, 495 (2000), hep-ph/0003030; 
K.A.~Milton, I.L.~Solovtsov, and O.P.~Solovtsova, Phys. Rev. D 
{\bf 65}, 076009 (2002), hep-ph/0111197. 
\bibitem{Nester}
A.V.~Nesterenko,  Mod. Phys. Lett. A {\bf 15}, 2401
(2000); A.V.~Nesterenko,  Phys. Rev. D {\bf 64}, 116009 (2001);
A.V.~Nesterenko and I.L.~Solovtsov,  Mod. Phys. Lett. A {\bf 16},
2517 (2001).
\bibitem{Sidor}
A.V.~Sidorov, Nuovo Sim. A {\bf 112}, 1527 (1999), hep-ph/9810460.
\end{thebibliography}
\end{document}